\documentclass[preprint,12pt]{elsarticle}

\usepackage[T1]{fontenc}
\usepackage{todonotes}
\usepackage[normalem]{ulem}

\usepackage{soul}

\newskip\movedskip
\newcommand{\movespaceafter}[1]{%
    \movedskip=0pt%
    \ifhmode\ifdim\lastskip=0pt\else\movedskip=\lastskip\unskip\fi\fi
    #1\ifdim\movedskip=0pt\else\hskip\movedskip\fi
    \ignorespaces}

\definecolor{mossgreen}{HTML}{146614}

\newcommand{\DONE}[2][]{\sout{#2} \checkmark\ \ifthenelse{\equal{#1}{}}{}{#1: }Done}

\usepackage{amssymb}
\usepackage{natbib}
\usepackage{booktabs}
\usepackage{amsfonts}
\usepackage{graphicx}
\usepackage{listings}
\usepackage{subcaption}
\usepackage{tabularx}
\usepackage{breqn}
\usepackage{acro}
\usepackage{listings}
\usepackage{ragged2e}
\usepackage{multirow}
\usepackage{makecell}
\usepackage{wrapfig}

\lstdefinelanguage{json}{
    numberstyle=\small,
    rulecolor=\color{black},
    showspaces=false,
    showtabs=false,
    breaklines=true,
    breakatwhitespace=true,
    basicstyle=\fontsize{7}{7}\ttfamily,
    upquote=true,
    string=[b]"
}

\lstdefinelanguage{turtle}{
    numberstyle=\small,
    rulecolor=\color{black},
    showspaces=false,
    showtabs=false,
    breaklines=true,
    breakatwhitespace=true,
    basicstyle=\fontsize{7}{7}\ttfamily,
    upquote=true,
    string=[b]",
    keywordstyle=\bfseries,
    keywords={rdf,type,xsd,double}
}

\newcommand{\dash}{\kern -.07em\_\kern .07em}
\newcommand{\NL}{\hfil\null\penalty-9999}

\usepackage{amssymb}
\usepackage{pifont}
\newcommand{\cmark}{\ding{51}}%
\newcommand{\xmark}{\ding{55}}%

\DeclareAcronym{EMBER}{short=EMBER,long=Elastic Malware Benchmark for Empowering Researchers}
\DeclareAcronym{SOREL}{short=SoReL,long=Sophos\slash ReversingLabs 20 million sample dataset}
\DeclareAcronym{MAEC}{short=MAEC,long=Malware Attribute Enumeration and Characterization}
\DeclareAcronym{PARCEL}{short=PARCEL,long=Parallel Class Expression Learner}
\DeclareAcronym{MSMCC}{short=MSMCC,long=Microsoft Malware Classification Challenge}

\usepackage[inline]{enumitem}

\journal{Computers \& Security}

\usepackage{amsmath}

\begin{document}

\def\ocel{OCEL}
\def\celoe{CELOE}
\def\parcel{PARCEL}
\def\spacel{SPACEL}
\def\parcelex{\spacel}

\def\nN{\ensuremath{\mathbb{N}}}
\def\nZ{\ensuremath{\mathbb{Z}}}

\newcommand{\numr}[2]{\mathop{{#1}#2}}
\newcommand{\atleast}[1]{\numr{\geqslant}{#1}}
\newcommand{\atmost}[1]{\numr{\leqslant}{#1}}
\newcommand{\exactly}[1]{\numr{=}{#1}}

\def\lang{\ensuremath{\mathcal{L}}}
\def\kb{\ensuremath{\mathcal{K}}}
\def\interp{\ensuremath{\mathcal{I}}}

\def\cn{\ensuremath{N_C}}
\def\rn{\ensuremath{N_R}}
\def\dn{\ensuremath{N_I}}

\def\concepts{\ensuremath{\bf C}}
\def\roles{\ensuremath{\bf R}}

\def\tbox{\ensuremath{\mathcal{T}}}
\def\rbox{\ensuremath{\mathcal{R}}}
\def\abox{\ensuremath{\mathcal{A}}}

\def\entails{\ensuremath{\vDash}}
\def\nentails{\ensuremath{\nvDash}}

\makeatletter
\newcommand\footnoteref[1]{\protected@xdef\@thefnmark{\ref{#1}}\@footnotemark}
\makeatother

\begin{frontmatter}



\title{Semantic Data Representation for Explainable Windows Malware Detection Models
}


\author[inst1]{Peter \v{S}vec\corref{cor1}}

\author[inst1]{\v{S}tefan Balogh}

\affiliation[inst1]{organization={Institute of Computer Science and Mathematics, Faculty of Electrical Engineering and Information Technology, Slovak University of Technology}, 
            addressline={Ilkovi\v{c}ova~3}, 
            city={Bratislava},
            country={Slovakia}}

\author[inst2]{Martin Homola}
\author[inst2]{J{\'a}n K\v{l}uka}
\author[inst2]{Tom{\'a}\v{s} Bist{\'a}k}
\author[inst2]{Peter Anthony}

\affiliation[inst2]{organization={Department of Applied Informatics, Faculty of Mathematics, Physics and Informatics, Comenius University in Bratislava},
            addressline={Mlynsk\'{a} dolina}, 
            city={Bratislava},
            country={Slovakia}}

\begin{abstract}
    Ontologies are a standard tool for creating semantic schemata in many knowledge intensive domains of human interest. They are becoming increasingly important also in the areas that have been until very recently dominated by subsymbolic knowledge representation and machine-learning-based data processing. One such area is information security, and specifically, \emph{malware detection}.
    We thus propose \emph{PE Malware Ontology} that offers a reusable semantic schema for Portable Executable (PE -- the Windows binary format) malware files. This ontology is inspired by the structure of the dataset \acf{EMBER}, which focuses on the static malware analysis of PE files. With this proposal, we hope to provide a unified semantic representation for the existing and future PE-malware datasets and facilitate the application of symbolic, neuro-symbolic, or otherwise explainable approaches in the PE-malware detection domain, which may produce interpretable results described by the terms defined in our ontology. In addition, we also publish semantically treated \acs{EMBER} data, including fractional datasets, to support the reproducibility of experiments on \acs{EMBER}.
    We supplement our work with a preliminary case study, conducted using concept learning, to show the general feasibility of our approach. While we were not able to match the precision of the state-of-the-art machine learning tools, the learned malware discriminators were interesting and highly interpretable.
\end{abstract}




\end{frontmatter}


\section{Introduction}
\label{sec:introduction}

\subsection{Background}
The application of machine learning (ML) in the area of malware detection is becoming more and more popular \citep{UCCI2019123,Pramanik2019,gibert2020rise,shaukat2020survey,Supriya2020malware,SinghS21,RaffFZAFM21,Aggarwal2021,TayyabKDKL22}.
While ML classifiers can identify malware instances to large success, most of them are not able to output any form of \emph{interpretable} justification for why a particular instance is classified as malware. The malware research community has already recognized this issue, and works trying to employ various explainable methods have recently started appearing. Among the exploited techniques, we may find ex-post explanation methods \cite{Amich2021} (such as LIME \cite{LIME} and SHAP \cite{SHAP}); methods adopted from explainable image classification \citep{Iadarola2021,Marais2022}; and methods for learning decision trees \cite{MojisKenyeres2023}, random forests \citep{Mills2019}, decision lists \cite{Dolejs2022}, and concepts \cite{svec2021experimental}, which produce symbolic class characterizations of different kinds and hence loosely fall under Structured Machine Learning (SML) \cite{westphal2019sml}.

However, Wissner-Gross observed that today, ``datasets -- not algorithms -- might be the key limiting factor to the development of artificial intelligence'' \citep{wissner-gross}. Indeed, the shift towards artificial-intelligence (AI) toolkits in malware analysis was, to a large extent, enabled by the availability of 
datasets suitable to train such
malware-detection models.
\acf{EMBER} \citep{anderson2018ember}
(approx.\ 1.1 million samples), \acf{SOREL}
\citep{harang2020sorel} (approx.\ 20 million samples), and BODMAS \cite{yang2021bodmas} (approx.\ 134 thousand samples) are among the most popular public datasets used for this purpose.
They both cover the domain of PE malware (i.e., Windows executables and libraries) and contain data resulting from static malware analysis \cite{sikorski2012practical}. The advantage of these datasets is that they are already widely analyzed by
security experts and extensively used%
\footnote{\label{footnote1}According to Google Scholar, EMBER \cite{anderson2018ember} had 700 citations, BODMAS \cite{yang2021bodmas} had 190, and \ac{SOREL} \cite{harang2020sorel} had 139, at the time of the last revision of this paper.}
in scientific studies.
They also provide extensive information on static features in a structured textual format for each sample.
%
%
%

\subsection{Motivation}
\looseness=-1
This data is essentially symbolic -- the textual characterization of each sample can
be readily interpreted as a set of meaningful properties -- and it can be viewed as (or exported into) a knowledge base or a knowledge graph \cite{knowledge-graph}. 
Such a treatment of the data would enable capitalizing on
knowledge-enabled tools or even tools rooted in the more recent
neuro-symbolic AI movement \cite{NSAI}, with the outlook of improving the interpretability of the resulting classifiers.
The missing element is a unified semantic schema (i.e., an \emph{ontology} \cite{ontology}) that would further allow us to (i) seamlessly integrate data regardless of its source, (ii) inter-relate outputs resulting from different tools, and (iii) develop a unified approach for their interpretation in human language \citep{AndroutsopoulosLG13,CimianoLNU13,GalanisA07}. Several ontologies \cite{Sikos2019} have already been developed for the cyber-security domain, and even for malware detection \cite{XiaDJZ17,DingWZ19}, but to our best knowledge, there is no suitable ontology that fits EMBER, \ac{SOREL}, and other similar datasets.

To give an example, SML approaches, such as \emph{concept learning} \cite{DL-Learner}, inductive logic programming \cite{HocquetteNJC24,ILP}, or decision-tree learning \cite{MojisKenyeres2023} may be used in combination with a semantically annotated malware dataset to obtain symbolic characterizations of malware samples in the form of concepts or other symbolic expressions \citep{svec2021experimental}.
\emph{Knowledge-base embedding} \citep{TRANSE} may be used to improve the efficiency of subsymbolic models by injecting prior symbolic knowledge about samples into their subsymbolic representations. More recently, \emph{recognition of neural-network activation patterns} and their alignment with ontology concepts \cite{deSousaRibeiro2021} may be used to explain and diagnose the trained neural-network classifiers.

Another aspect one must handle when working with \ac{EMBER} and \ac{SOREL} is their enormous size -- they both contain millions of samples.
Some of the studies using \ac{EMBER} and \ac{SOREL} \citep{vinayakumar2019,liu2020,ghouti2020} had to resort to reducing the size of the
dataset of interest, especially if training the employed classifier was computationally demanding. However, no unified methodology for
reducing the datasets has been established.

\subsection{Goals of this study}
In this work, we intend to alleviate the applicability of ML-based methods, especially interpretable and  explainable methods and neuro-symbolic methods, on the popular malware datasets by setting the following three goals:
\begin{description}
\item[Unified semantic representation:] To provide a unified semantic schema for the domain of PE malware that may be then used to represent any relevant data sample regardless of its source.
\item[Interpretability of results:] To ensure that the schema provides a suitable vocabulary that is meaningful and recognized by human users, making prototype samples, concept descriptions, rules, or any other characterizations expressed in the vocabulary understandable and explainable.
\item[Reproducibility and comparability of experiments:] To ensure that experiments may be
executed on datasets of different suitable sizes without the need to reduce
the full EMBER dataset to smaller ones repeatedly in each study, which would enable a direct comparison of the outcomes of the experiments not only in numeric measures (e.g., precision and accuracy) but also in terms of the constructed explanations.
\end{description}

\subsection{Contribution}
To address these goals, we have developed a unified ontology that
can provide a reusable semantic schema for PE malware files.  To the best
of our knowledge, such an ontology was missing so far.
We have also aligned the nomenclature used to describe actions possibly performed by PE files with the
\acf{MAEC} standard \citep{maec}, which is widely accepted for malware descriptions.

Our goal was to improve the interpretability of the captured data. 
The ontology is partly based on the \ac{EMBER} dataset structure 
derived from the \ac{EMBER} data features, but it is not a direct
ontological mapping of this or any other individual dataset.
The emphasis is placed on data
features that are meaningful in malware characterization.
For this reason, we included only those features that make sense from an expert's point of view, while other, less meaningful features were left out. Some of the excluded features (e.g., file size) may even cause the trained classifiers to be prone to trial counterattacks.
Also, some of the features captured in the ontology can be considered \emph{derived}, in the sense that they combine multiple low-level EMBER data entries into a single meaningful feature. Two features are based on thresholds for
numeric features (e.g., ``high entropy''), which is useful for systems that have difficulties with handling numeric values.

But our goal was also to produce a dataset that could be effectively used as a standard benchmark to help advance learning algorithms in the malware detection domain and that would also offer unambiguous reproducibility of experimental results. To this end, we release \ac{EMBER} data (2018 version) translated into RDF semantic format, including smaller fractional datasets for the sake of a better comparison of diverse experiments that require smaller data volume than the full \ac{EMBER}.

Our ontology, together with the fractional datasets and the \ac{EMBER}\slash\ac{SOREL}-to-RDF mapping scripts, is available in our GitHub repository\footnote{https://github.com/orbis-security/pe-malware-ontology}.

To demonstrate the basic workflow with ontologically treated data and to show the general feasibility of the approach, we complement this report with a preliminary case study with DL-Learner \cite{DL-Learner}, arguably one of the most popular concept-learning tools today. The results have shown the general applicability of this approach. The achieved numeric precision measures could not yet match the best ML-based results from the literature, but this was not our goal in this study. On the other hand, we were able to demonstrate that DL-learner produced relevant concept expressions that show a high level of interpretability according to security experts involved in the study. This highlights the importance of the introduced ontology as it was achieved mainly thanks to the selection of relevant properties to be captured in it in form of a meaningful and well-understandable vocabulary from which these expressions were built.

All in all, this work contributes to the improvement of handling the data in malware analysis and to the improvement of the applicability of interpretable and XAI methods in this area. Such development is much in line with the European AI Act and other current regulation and research trends, which put forward the issue of the trustworthiness of AI tools applied especially in critical areas such as cybersecurity. Finally, our work also results in a clearly defined and well-motivated use case and a tailored semantic dataset with real-world data that can be instrumental in further development of SML and neuro-symbolic learning tools which extends its significance in other application areas where such tools may be applied.

\subsection{Paper Structure}
The rest of the paper is structured as follows: Section~\ref{sec:prelim} provides the necessary preliminaries on ontologies and description logics; Section~\ref{sec:dataset} reviews the data sources on top of which this work was built; Section~\ref{sec:preprocessing} explains how the data were pre-processed and enriched; Section~\ref{sec:ontology} describes the PE Malware Ontology; Section~\ref{sec:fractions} describes the exported semantic datasets; Section~\ref{sec:case-study} contains a proof-of-concept case study that shows the viability of the overall approach; we reflect on the related works in Section~\ref{sec:related}; and we conclude with a discussion in Section~\ref{sec:conclusions}.

\section{Preliminaries}
\label{sec:prelim}

Ontologies were successfully applied to establish shared, unambiguous, and potentially reusable semantics for data in areas such as biology and medicine \cite{Stearns2001, Bodenreider2006,Ivanovic2014}, industry \citep{optiqueVQS}, e-commerce \citep{Hepp08,Tamma2005}, web syndication \citep{Guha2016}, e-learning \citep{al-yahya2015ontologies,HomolaKKMC19,HomolaKUK23}, and many others. Their significance in computer security is also on the rise \citep{bromander2016semantic,mavroeidis2017cyber,syed2016uco,ulicny2014inference}.   

An ontology is defined as a \emph{formal conceptualization of a domain of interest} \cite{gruber1993,ontology}. Ontologies thus define the relevant terms in a given domain and their relations using some suitable and unambiguous language. Among the most popular languages are description logics (DLs) \citep{dlhb} or their variant known as Web Ontology Language (OWL) \citep{OWL1,OWL2}. In these languages, ontologies are built from atomic entities, such as individuals (particular objects in the domain), classes (also concepts, which correspond to types of individuals), object properties (relations between individuals), and data properties (attributes used to assign data values to individuals).

To provide an example, in the PE malware domain, we will encounter individuals corresponding to actual sampled files with their unique names derived from hash sums (e.g.\ \texttt{aba129a3d1ba9d307dad05617f66d8e7}) but also distinguished prototype individuals e.g.\ corresponding to possibly executed actions (\texttt{create-window}, \texttt{sleep-process}, etc.); classes such as  \texttt{PEFile},\linebreak[4] \texttt{ExecutableFile}, \texttt{DynamicLinkLibrary}; object properties such as \texttt{has\_file\_}\linebreak[1]\texttt{feature} or \texttt{has\_}\linebreak[1]\texttt{section}; and data properties such as \texttt{section\_entropy} or \texttt{imports\_count}.
For the purpose of this section, we will not be concerned by the exact meaning of these symbols (it will be given later on). Note the widely accepted naming convention: classes start with upper case letters, while individuals and properties with lower case letters.

One of the main advantages of DLs is that we can construct complex classes, also called concept descriptions \cite{dlhb}, from the named classes of the ontology
(such as \texttt{DynamicLinkLibrary} or \texttt{NonexecutableEntryPoint}
in the PE Malware Ontology) in these formalisms
using DL concept constructors, such as complement~$\lnot$,
conjunction~$\sqcap$, and disjunction~$\sqcup$;
qualified existential~$\exists$, universal~$\forall$,
and number $\leqslant$, $=$,~$\geqslant$ restrictions on properties;
and nominals describing classes by enumerating all contained individuals 
(e.g., $\{$\texttt{create-window}$,\allowbreak{}$\texttt{sleep-process}$\}$).
For example, the class described by
\[\text{\small$\begin{aligned}
  \texttt{DynamicLinkLibrary}
  &\sqcap \exists\texttt{has\_action}.\,
        \lnot\{ \texttt{create-window}, \texttt{sleep-process} \}\\
  &\sqcap \forall\texttt{has\_file\_feature}.
        (\texttt{CLR} \sqcup \texttt{RegistryStrings})
\end{aligned}$}\]
refers to DLLs which can perform actions other than window creation
and process sleeping,
and the only distinguished file features they have (if any)
are the presence of the Common Language Runtime data directory
or some registry strings.
The values of object properties can be qualified by nested concept descriptions
and the values of data properties by XML Schema datatypes (e.g., \texttt{xsl:integer}) \cite{peterson2012datatypes},
optionally restricted by predicates (e.g., $[> 0]$ for positive numbers). For an illustration of this feature, consider the description
\[\text{\small$\begin{aligned}
    {\geqslant}2\,\texttt{has\_section}.(
    &\texttt{DataSection} \sqcap
    \exists\texttt{has\_section\_flag}.\,\texttt{Executable} \\
    &\sqcap \exists\texttt{section\_entropy}.\,\texttt{xsd:double}[\geq 4.5]),
\end{aligned}$}\]
which matches individuals that have at least two sections,
each of which is simultaneously a data section, has the executable flag set,
and its content entropy is a double-precision floating-point number
greater than or equal to~4.5.

In an ontology, complex concept descriptions may be used in axioms to define classes (or less strictly, to provide necessary or sufficient conditions of other classes). Such an axiom is called a general concept inclusion (GCI) and has the form $C \sqsubseteq D$, which intuitively means that all instances of a class $C$ are also instances of a class $D$, with both $C$ and $D$ being possibly complex descriptions. A collection of GCIs is called a TBox. An ontology (knowledge base) is formed by a TBox and an ABox, where the latter contains data, expressed by axioms of the form $C(a)$ and $P(a,b)$, with the meaning that the individual $a$ is an instance of the class $C$ and that the property $P$ connects the individuals $a$ and~$b$ (or the individual~$a$ with the value~$b$, in case $P$~is a data property). These axioms are, respectively, called class assertions and property assertions.

DLs provide ontologies with model-theoretic semantics, but we will omit the details (see e.g.\ Baader et al. \cite{dlhb}). It suffices to say that the semantics allows us to check if an ontology is consistent, where the complex GCI axioms of the ontology act as constraints. For instance, if a certain individual is assigned to both \texttt{PEFile} and \texttt{Section} in the ABox, the ontology will be inconsistent if it contains the respective disjointness constraint (\texttt{PEFile} ${}\sqsubseteq \neg$ \texttt{Section}), as intuitively, a section (a part of a file) is not a file.
The semantics also enables us to derive implicit knowledge, that is, valid statements which are not directly present in the ontology but are implied: if a statement $\phi$ (either a GCI or an ABox axiom) is valid (also entailed) in an ontology $\mathcal{O}$, this is denoted by $\mathcal{O}\models\phi$. For example, if an individual corresponding to some sample file is in the class \texttt{DynamicLinkLibrary}, it is entailed that it is also in all its superclasses, e.g.\ in \texttt{PEFile}.

Working with ontologies is well supported by many available software tools.
The computational tasks of consistency and entailment checking are readily implemented in a number of DL reasoners (e.g.\ \citep{hermit,konclude,elk}) which are well integrated into ontology development tools \citep{protege}. Also, properly designed ontologies show in general good levels of extensibility and they can be extended or combined with ontologies covering different domains as required by an application \cite{modular-ontologies,blomqvist-XD}.

\section{Methodology overview}
\begin{figure}
    \centering
    \includegraphics[width=\textwidth]{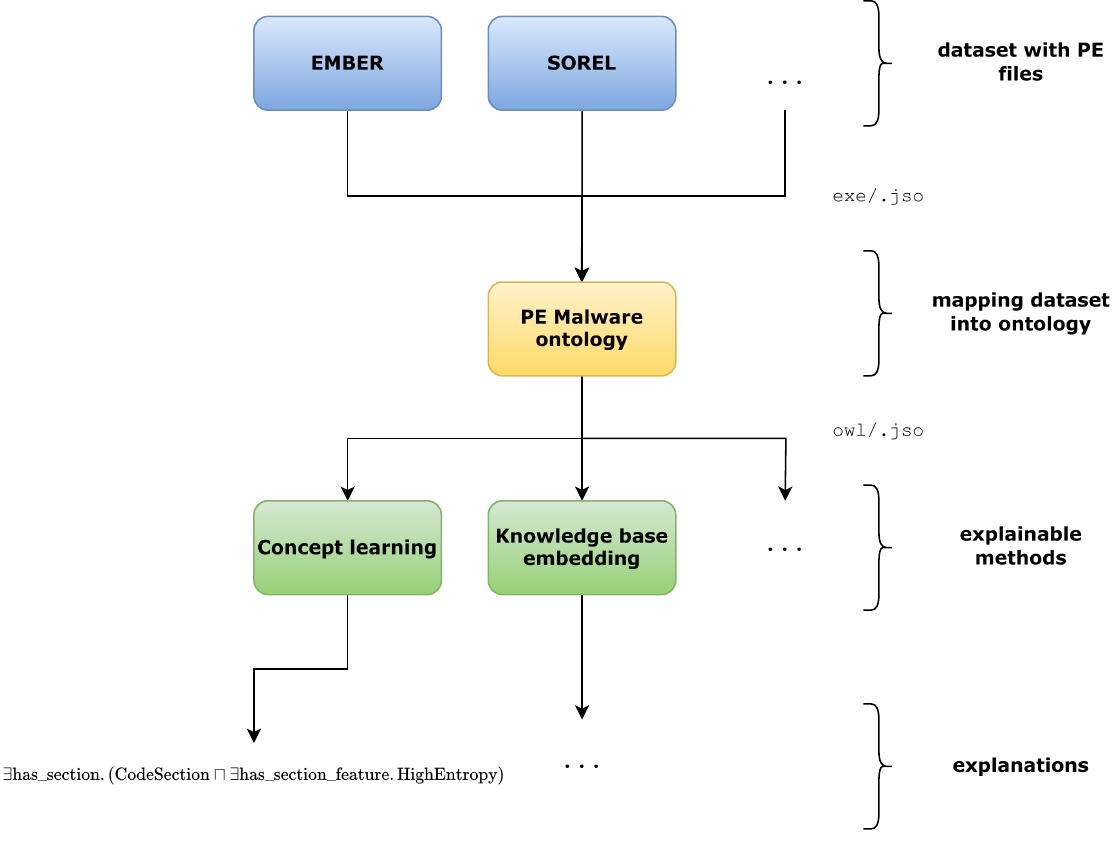}
    \caption{Overview of our methodology}
    \label{fig:scheme}
\end{figure}

One of the important aspects of our ontology is the fact that it is not limited to use only with the \ac{EMBER} dataset and concept learning algorithms. The overview of our methodology is depicted in Figure \ref{fig:scheme}. The scheme consists of the following elements:

\begin{description}
    \item[Dataset:] The foundation of the entire methodology is a dataset with PE executable files. While we used the \ac{EMBER} dataset in our experiments, our ontology can be used with any dataset that contains executable PE files or extracted static features (as in the case of \ac{EMBER}).

    \item[Ontology-based representation:] The second step in the workflow is the transformation of the dataset into the semantic format OWL using the ontology. Our script can be used to map the dataset from a specific JSON format. This format is used by the datasets \ac{EMBER} and \ac{SOREL}. The authors of the \ac{EMBER} dataset, together with the data itself, also published a script that can transform any PE file into this specific JSON format. Therefore, our ontology can be also applied to any dataset containing binary PE files.

    \item[Learning:] The result of the previous step is an ontology represented in OWL and populated with individuals, together with the specification of positive and negative examples in a JSON file. This dataset can be subsequently used in the learning process. In our work, we focused on concept learning (specifically, on the algorithms available in the DL-Learner framework), but these datasets can also be used with other methods that can work with knowledge bases (e.g.\ SML methods \cite{westphal2019sml} or nowledge-base embedding methods \cite{KBEmbeddingSurvey}).

    \item[Explanations:] The output of the whole process is a set of class expressions\slash rules (depending on the method used). This output offers not only a classification of the input sample, but also explanations that can be easily understood.
\end{description}

\section{Data sources}
\label{sec:dataset}

In this section, we describe in more detail the \ac{EMBER} dataset, which we used as the foundation for our ontology, and the modifications we proposed. Another very similar dataset is \ac{SOREL}, which contains more samples and it also includes disarmed malware binaries. However, the static properties of samples that these datasets offer are almost identical (a few are missing in \ac{SOREL}).

\subsection{\Ac{EMBER}}

The \acf{EMBER} dataset \citep{anderson2018ember} contains a total of 1.1 million samples and includes 400,000 (400\,k) malicious samples, 400\,k benign samples, and 300\,k unlabeled samples, so the dataset can also be used for unsupervised learning. The labeled samples from the dataset are also divided into training and testing sets (600\,k for training and 200\,k for testing). The dataset is composed of a collection of JSON objects, where each object represents data statically extracted from the PE file of one sample. A simplified example of one sample can be seen in Listing~\ref{lst:ember}. The static properties themselves are organized as follows:

\begin{lstlisting}[float=tbhp, language=json, caption={Static features for single binary sample.}, label={lst:ember}]
{
  "sha256": "eb87d82ad7bdc1b753bf91858d2986063ebd8aabeb8e7e91c0c78db21982a0d6", 
  "md5": "aba129a3d1ba9d307dad05617f66d8e7", 
  "appeared": "2018-01",
  "label": 1,
  "avclass": "fareit",
  "histogram": [ 96506, 8328, 5582, ... ],
  "byteentropy": [0, 4229, 269, 247, ... ],
  "strings": {
    "numstrings": 7762,
    "avlength": 181.60641587219789,
    "printabledist": [591, 51, 96, 46, ... ],
    "printables": 1409629,
    "entropy": 5.037064474164528,
    "paths": 0,
    "urls": 9, 
    "registry": 0,
    "MZ": 11
  }, 
  "general": {
    "size": 2261028, 
    "vsize": 1912832, 
    "has_debug": 0, 
    "exports": 0, 
    "imports": 17,
    "has_relocations": 1, 
    "has_resources": 1, 
    "has_signature": 0, 
    "has_tls": 1,
    "symbols": 0
  },
  "header": { 
    "coff":{
      "timestamp": 708992537,
      "machine": "1386",
      "characteristics": ["CHARA_32BIT_MACHINE", "BYTES_REVERSED_LO", "EXECUTABLE_IMAGE", ... ]
    },
    "optional": {
      "subsystem": "WINDOWS_GUI", 
      "dll_characteristics": [], 
      "magic": "PE32", 
      "major_image_version": 0, 
      "minor_image_version": 0, 
      "major_linker_version": 2, 
      "minor_linker_version": 25,
      ...
    }
  },
  "section": { 
    "entry": "CODE", 
    "sections": [
      {
        "name": "CODE",
        "size": 443392,
        "entropy": 6.532932639432919,
        "vsize": 442984,
        "props": ["CNT_CODE", "MEM_EXECUTE", "MEM_READ"]
      },
    ...
    ]
  },
  "imports": {
    "kernel32.dll": ["DeleteCriticalSection", "TlsSetValue", "Sleep", ... ],
  },
  "exports": [],
  "datadirectories": [ { "name": "EXPORT_TABLE", "virtual_address": 0 }, ... ]
}
\end{lstlisting}

\begin{description}
	\item[General file information:] This set of features is dedicated to general information about the file, such as the file's size, the respective numbers of the imported/exported functions, and the indicators of whether the file contains a digital signature, debugging symbols, a TLS section, etc.
	\item[Header information:] These are the properties found in the file's headers, such as the target architecture for which the file was compiled, the linker version, various timestamps, etc.
	\item[Section information:] This set is dedicated to individual sections in the binary file. For each section, there is its name, content type (code, initialized or uninitialized data), various properties (such as read, write, and execute rights), and the value of the entropy of the section's content.
	\item[Imported functions:] The list of imported functions organized by the DLL (here, it is necessary to note that if the file imports a certain function, it does not necessarily have to be called in the code).
	\item[Exported functions:] The list of exported functions. These are mostly included only in cases where the PE file is a library.
\end{description}

In addition to the categories mentioned above, the \ac{EMBER} dataset also contains other numerical properties such as byte histogram, byte-entropy histogram, or simple statistics for the strings found in the file.

\subsection{Other sources}
\label{sec::othersources}

\begin{table}[htbp]\scriptsize
	\caption{Public malware datasets.}
	\centering
    \setlength{\tabcolsep}{.35em}
	\begin{tabular}{l@{}rrrrlll}
		\toprule
		\textbf{Name} & \textbf{Samples} & \textbf{Malware} & \textbf{Benign} & \textbf{Unlabl'd} & \textbf{Platform} & \textbf{Features} & \textbf{Format} \\
		\midrule
 		\ac{EMBER} \cite{anderson2018ember} & 1,100\,k & 400\,k & 400\,k & 300\,k & Windows & static & JSON \\
        \ac{SOREL} \cite{harang2020sorel} & 20,000\,k & 10,000\,k & 10,000\,k & 0 & Windows & static & LMDB+binary \\
        MSMCC \cite{ronen2018microsoft} & 20\,k & 20\,k & 0 & 0 & Windows & static & disassembly \\
        MALREC \cite{severi2018m} & 66\,k & 66\,k & 0 & 0 & Windows & dynamic & PANDA \\
        Mal-API-2019 \cite{catak2019benchmark} & 7.1\,k & 7.1\,k & 0 & 0 & Windows & dynamic & CSV \\
        AVAST-CTU \cite{bosansky2022avast} & 49\,k & 49\,k & 0 & 0 & Windows & stat.+dyn. & JSON\\
        ClaMP \cite{clamp} & 5.2\,k & 2.7\,k & 2.5\,k & 0 & Windows & static & CSV \\
        MalImg \cite{nataraj2011malware} & 9.5\,k & 9.5\,k & 0 & 0 & Windows & static & greyscale image\\
        DREBIN \cite{arp2014drebin} & 129\,k & 5.6\,k & 123\,k & 0 & Android & static & text file \\
        MOTIF \cite{joyce2023motif} & 3.0\,k & 3.0\,k & 0 & 0 & Windows & static & JSON+binary\\
        Malpedia \cite{plohmann2017malpedia} & 7.8\,k & 7.8\,k & 0 & 0 & Multiple & stat.+dyn. & memory dumps \\
        MalDozer \cite{karbab2018maldozer} & 71\,k & 33\,k & 38\,k & 0 & Android & static & unknown \\
        BODMAS \cite{yang2021bodmas} & 134\,k & 57\,k & 77\,k & 0 & Windows & static & JSON  \\
        \bottomrule
	\end{tabular}
	\label{tab:publicdatasets}
\end{table}

Inspired by the success of \ac{EMBER}, Harang and Rudd published \acf{SOREL} \citep{harang2020sorel} to scale up the volume of the available samples.
\ac{SOREL} is structured similarly to \ac{EMBER} and covers the same static features with very minor differences. The \ac{SOREL} samples are again provided in the form of JSON objects.
In addition to approx.\ 20 million (20\,M) labeled JSON-structured samples, \ac{SOREL} also contains approx.\ 10\,M real malware binaries that have been disarmed by modifying their header files. Despite being based on \ac{EMBER}, the ontology presented in this work may be directly applied to \ac{SOREL} data as well. In fact, our ontology is compatible with any dataset that contains raw PE files which the \ac{EMBER} translation script can translate into JSONs (which can be subsequently translated into our semantic representation). Another similar dataset is BODMAS \cite{yang2021bodmas}, which contains 134\,k samples with same features as \ac{EMBER} and \ac{SOREL} (i.e. it can be directly used with our ontology).

Our semantic representation focuses solely on static features (i.e., the features that are extracted from a PE file without its execution). Some datasets are also utilizing dynamic features (extracted from runtime). While these features can be useful for malware detection, they are more difficult to extract and most of the state-of-the-art datasets are using only the static features. Table \ref{tab:publicdatasets} summarizes the attributes of all major public malware datasets. For each dataset, we list the number of samples it contains, the platform these samples are targeted at, the type(s) of features included, and the format in which the data are stored.

As we can see in Table \ref{tab:publicdatasets}, there are currently no malware datasets that use a semantic format. Among the most popular sources (that provide real binaries) are {VirusTotal} \citep{virustotal}, which is, however, a paid service (along with benign files), {VirusShare} \citep{virusshare}, which currently contains approx.\ 50\,M malicious samples, and {MalShare} \citep{malshare}. \ac{MSMCC} \cite{ronen2018microsoft}, which is a dataset released by Microsoft, contains 20\,k samples from 9 different families in the form of a disassembly (without header). Known datasets that contain dynamic properties include \textsc{Malrec} \citep{severi2018m}, which captures the overall activity of malware by logging all sources of indeterminism in the system, such as system calls, peripheral devices, etc. In total, it contains approx.\ 66\,k samples.
{Mal-API-2019} \citep{catak2019benchmark} is a smaller dynamic dataset that focuses primarily on logging API calls and contains approx.\ 7\,k~samples. In 2022, a dataset from AVAST was released \citep{bosansky2022avast}, which, unlike previous works, combines static and dynamic features and contains approx.\ 50\,k malware samples (its main purpose is to classify malware into individual families). Other well-known datasets are {ClaMP} \citep{clamp}, which contains only Portable Executable (PE) file headers (approx.\ 5\,k), and {MalImg} \citep{nataraj2011malware}, which contains approx.\ 10\,k malicious samples from different families encoded as grayscale images. The {DREBIN} \citep{arp2014drebin} and MalDozer \cite{karbab2018maldozer} datasets are also worth mentioning, but unlike the previous works, they contain binary samples from the Android platform. \textsc{MOTIF} \cite{joyce2023motif} is a small-scale dataset that focuses on the ground-truth labels for the provided samples.


\section{Dataset standardization and pre-processing}
\label{sec:preprocessing}

In this section we introduce details regarding the pre-processing of the EMBER dataset and its standardization. Main goal of the pre-processing stage was to create various derived features from the original dataset, that would improve the interpretability of learned expressions (these features were created either as combination of multiple low-level features from EMBER dataset or based on threshold values).  Standardization, on the other hand, included selecting a specific API calls from MAEC standard and mapping them into the ontology (this step also helped in reducing the large number of API calls from the EMBER dataset).

\subsection{Dataset standardization}
\label{sec:standardization}

Individual samples may import a large number of functions from various standard DLLs implementing system APIs. They provide useful clues about the possible actions a sample file can perform when running, although not all imported functions may be called and not all of these actions are relevant from the malware-detection perspective. In order to extract only relevant information from samples' imports to the ontological representation in a way that is aligned with standard best practice, we have turned to \acf{MAEC} \citep{maec}. \acs{MAEC} is a community-developed structured language for describing information about malware, combining static and dynamic features (different kinds of behavior, interactions between processes, and so on). We map the space of imported functions to \textit{malware actions}, defined in one of \acs{MAEC}'s vocabularies \citep{maec-vocabs}. Not only does this mapping aid the standardization of the dataset, but it also achieves an effect similar to dimensionality reduction in traditional ML~-- irrelevant imported functions are disregarded and multiple functions with a similar effect can be mapped to a single action. For instance, the \texttt{EnumProcesses}, \texttt{Process32First}, and \texttt{Process32Next} WIN32 API functions are all mapped to the \texttt{enumerate-processes} action. The complete mapping of API functions to actions is available in our GitHub repository.

In order to map more API calls, we have extended the MAEC actions vocabulary with additional actions summarized in Table~\ref{tab:extendedactions}. A few new actions pertaining to calls to cryptographic functions, commonly used by malware, have been added. Moreover, MAEC \texttt{send-http-\textit{method}-request} actions (\texttt{send-http-get-request}, \texttt{send-http-post-request}, etc.) have been all replaced with a more general \texttt{send-http-request} action for the following reason. The \acs{MAEC} vocabulary describes malware actions in general, i.e., regardless of whether they are determined by static or dynamic analysis. However, an import of (or a call to) the \texttt{HttpSendRequest} API function can be mapped to a specific \texttt{send-http-\textit{method}-request} action only if its input parameters are known, which is information not available in static-analysis datasets, such as \acs{EMBER}.

Furthermore, we have categorized the actions into several classes, such as networking, file access, or system manipulation, in order to aid generalization in concept-learning applications of our dataset. We describe these classes in more detail in Section~\ref{sec:actions}.

\begin{table}[htbp]\small
	\caption{Extensions of the MAEC actions vocabulary}
	\centering
	\begin{tabular}{ll}
		\toprule
		\textbf{Action} & \textbf{Description} \\
		\midrule
		\texttt{encrypt}
        & The action of file encryption \\
		\texttt{decrypt}
        & The action of file decryption \\
		\texttt{generate-key}
        & The action of cryptographic key generation \\
		\texttt{send-http-request}
        & The action of sending an HTTP client request to a server\\
		\bottomrule
	\end{tabular}
	\label{tab:extendedactions}
\end{table}

\subsection{File and section features}
\label{sec:computed-features}

Samples in the \ac{EMBER} dataset are characterized by a number of properties of various types (see Listing~\ref{lst:ember}). We have selected the most salient of them based on malware detection expert knowledge \cite{sikorski2012practical, kleymenov2022mastering} to be mapped to the ontological representation as \emph{features} of samples (PE files) or samples' sections. All features are listed as OWL classes later in Section \ref{sec:ontology}, in Tables \ref{tab:filefeatures} and~\ref{tab:sectionfeatures}.
Here, we explain how the features are obtained from the information available in the \ac{EMBER} dataset. In this respect, there are three kinds of such features which we call (i)~directly represented features, (ii)~pre-processed features, and (iii)~derived features.

The \emph{directly represented features}, such as \texttt{Relocations}, arise from boolean (actually, 0/1) sample properties such as \texttt{general.\allowbreak has\_\allowbreak relocations}. The original properties mapped to these features are listed within the description of features in Table~\ref{tab:filefeatures}.

The \emph{pre-processed features} are obtained by pre-processing the original sample or section properties that are not directly represented in the ontology. There are two features of this kind: The \texttt{CLR} feature signifies the presence of a non-empty Common Language Runtime data directory (in the \texttt{datadirectories} property) in the sample, which is indicative of .NET binaries. The \texttt{Nonexecutable\-Entry\-Point} feature indicates that the sample's entry point is located in a non-executable section, which is checked by inspecting the \texttt{props} of the member of the \texttt{section.sections} array whose \texttt{name} is equal to the \texttt{section.entry}. Machine-learning algorithms can thus take advantage of these two important features even though the underlying properties (\texttt{datadirectories}, \texttt{section.entry}) are not represented in the ontology directly.

The \emph{derived features} are obtained by pre-processing the sample and section properties that are also represented directly in the ontological dataset.
They are sufficiently simple to be defined by DL descriptions over these properties (such as ${\geqslant}2\,\texttt{has\_section}.\,\allowbreak\exists\texttt{has\_\allowbreak section\_\allowbreak flag}.\,\allowbreak\texttt{Executable}$ for the \texttt{Multiple\allowbreak Executable\allowbreak Sections} feature). However, these features are known to experts to be important malware indicators and are thus worth naming explicitly. Moreover, assigning these features to samples during dataset pre-processing exposes them even to tools that lack the required expressivity (cardinality restrictions, data type restrictions, enumerated data types with long lists of literals) or for which enabling it imposes a significant performance penalty. If, e.g., a classifier trained on the dataset is capable of working efficiently with constructs required to define some of the derived features, these may be considered redundant and the user of our dataset may instruct the tool to selectively ignore them so as not to enlarge the dimensionality or search space unnecessarily. Likewise, if the user targets a learning algorithm to learn from the derived features, e.g., for the sake of efficiency, it may be indicated to remove or ignore the original data from which they were constructed.

Two features derived from the numerical sample properties in the \ac{EMBER} dataset stand out due to their important role in malware detection and a less trivial way of determining a suitable threshold controlling the assignment of the feature to a sample. These features are \emph{a low number of imports} of the sample and \emph{a high entropy} of a sample's section. They indicate that the sample may be packed. The threshold values for these features are inspired by \textit{pestudio} \citep{pestudio}, a tool commonly used by security teams for the initial assessment of malware samples. Based on \textit{pestudio}'s defaults, a sample importing less than 10 functions is considered to have a low number of imports, and a section with an entropy greater than~7.0 is marked as having a high entropy.

\begin{figure}[htbp]
\begin{subfigure}{.5\textwidth}
  \centering
  \includegraphics[width=.95\linewidth]{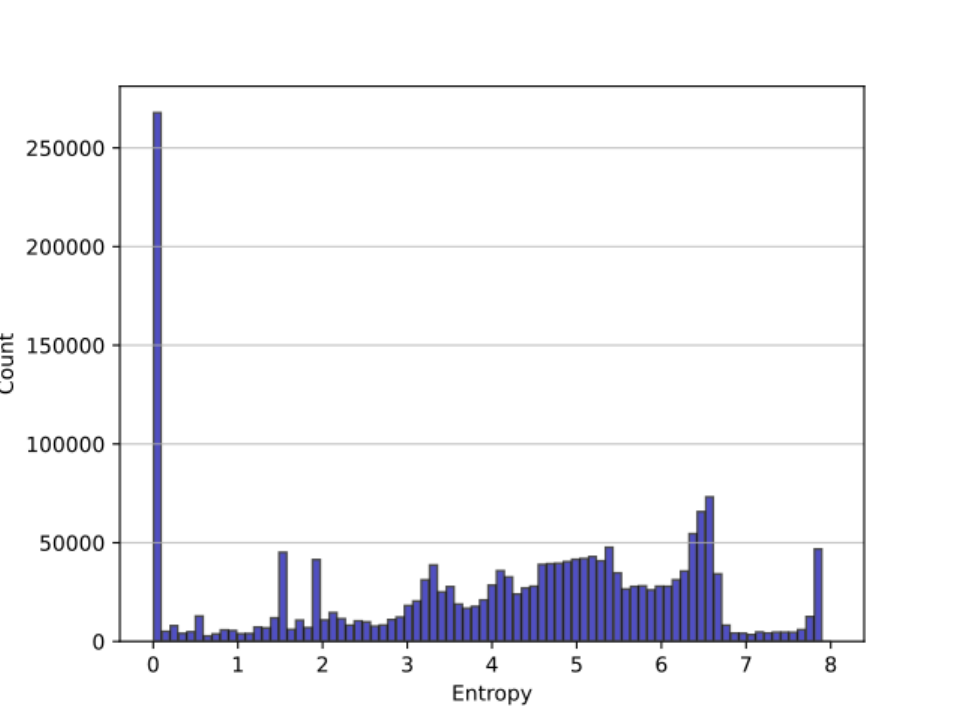}  
  \caption{Benign entropy histogram}
  \label{fig:hist-entropy-benign}
\end{subfigure}
\begin{subfigure}{.5\textwidth}
  \centering
  \includegraphics[width=.95\linewidth]{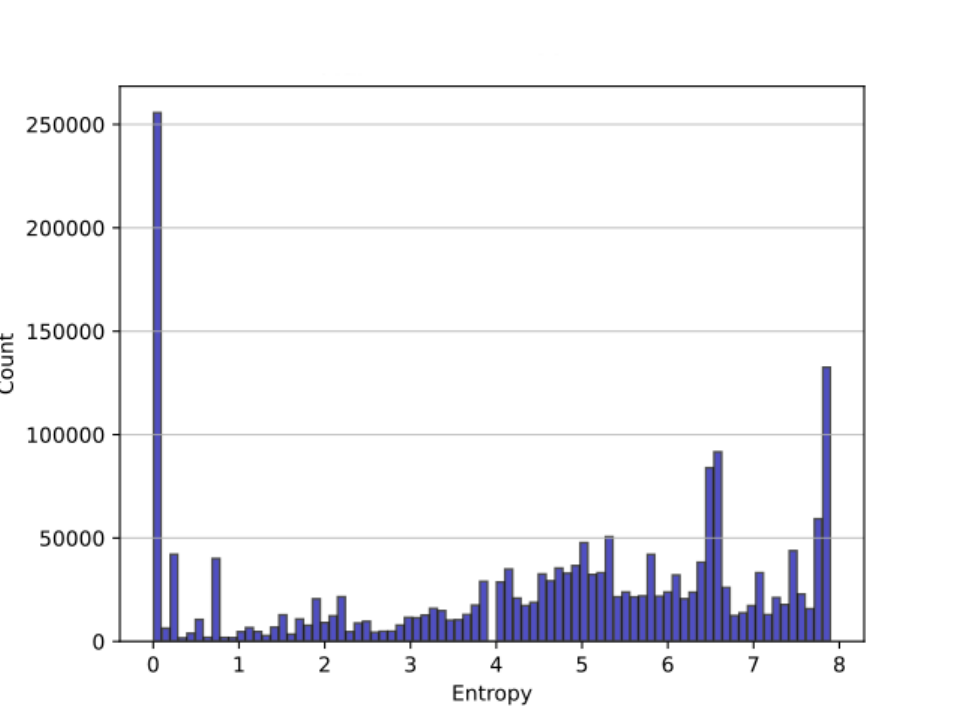}  
  \caption{Malware entropy histogram}
  \label{fig:hist-entropy-malware}
\end{subfigure}

\leavevmode

\begin{subfigure}{.5\textwidth}
  \centering
  \includegraphics[width=.95\linewidth]{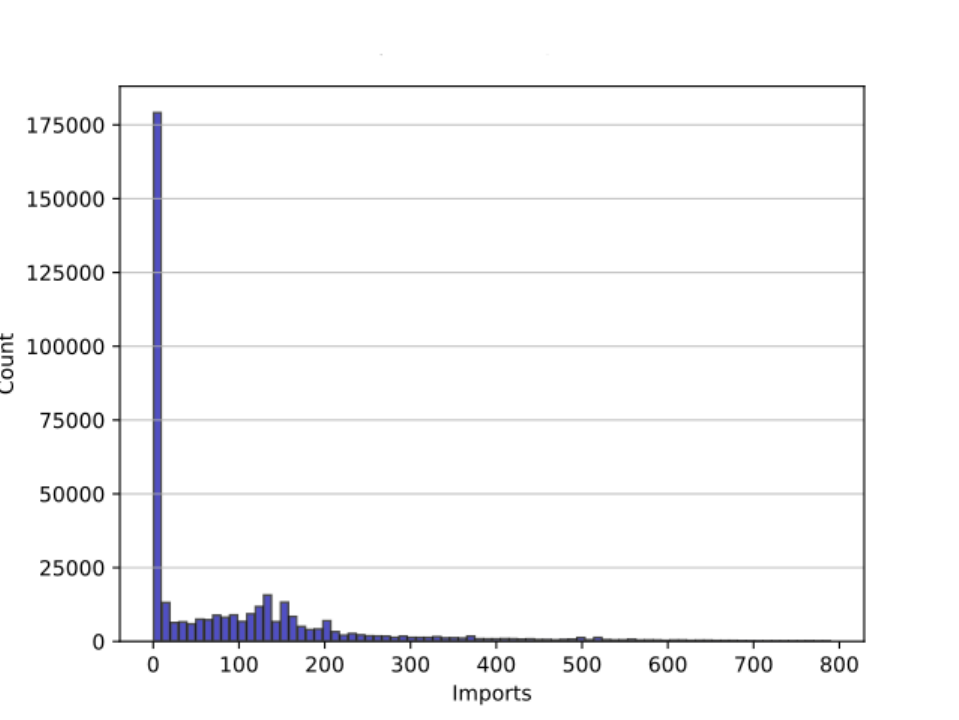}  
  \caption{Benign imports histogram}
  \label{fig:hist-imports-benign}
\end{subfigure}
\begin{subfigure}{.5\textwidth}
  \centering
  \includegraphics[width=.95\linewidth]{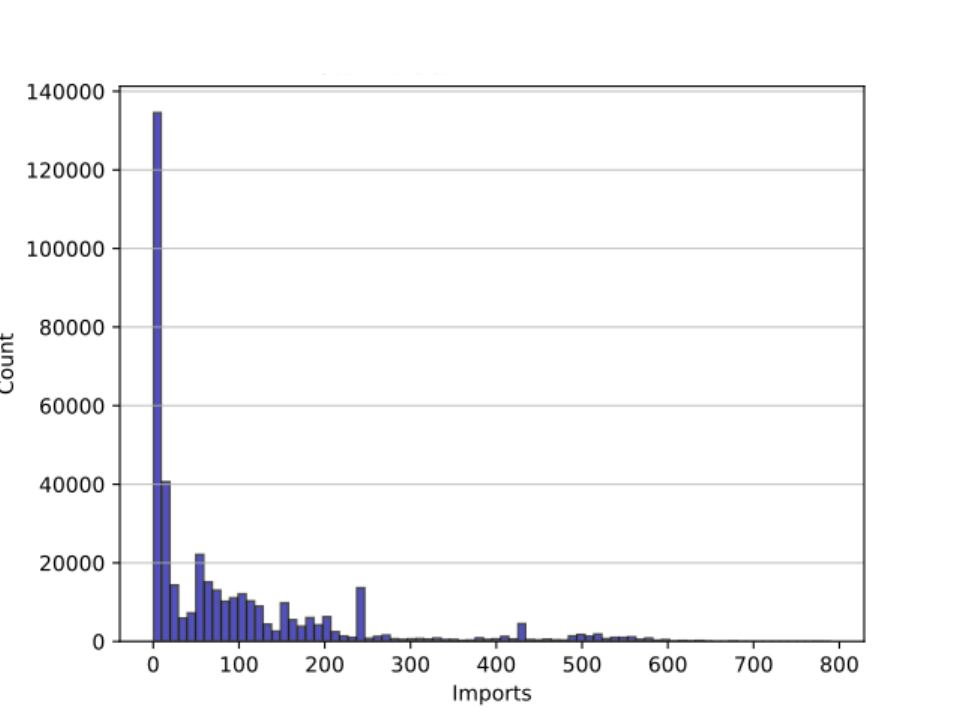}
  \caption{Malware imports histogram}
  \label{fig:hist-imports-malware}
\end{subfigure}
\caption{Entropy and imports histograms for the \ac{EMBER} dataset.}
\label{fig:histograms}
\end{figure}

In order to verify the threshold values from the \textit{pestudio} tool, we performed a basic statistical analysis for the entire EMBER dataset. Specifically, we were interested in histograms representing the number of imports for each sample and the entropy of each section in the dataset. The results are depicted in Figure~\ref{fig:histograms}. The threshold value for section entropy was confirmed. We can see that there are significantly more sections whose entropy is higher than~7.0 in the malware samples compared to the number of such sections in the benign samples.
As for the number of imports, the histograms in Figs.~\ref{fig:hist-imports-benign} and~\ref{fig:hist-imports-malware} do not show any clear indicative value that separates a significant volume of malware and benign samples. In particular, a large number of both kinds of samples import less than~10 API functions. This feature alone is thus not a strong indicator of a sample's maliciousness, although it may become useful in combination with other features. Hence, we decided to include it anyway and keep \emph{pestudio}'s preset threshold of~10.

Another non-trivially derived feature of sections is \emph{a non-standard name}. While we map section names to the ontological representation, describing this feature in OWL~2 requires a data property restriction with a negated list of section names usually produced by compilers and linkers. We consider it improbable that a learning algorithm would discover such a restriction. We have thus opted to collect usual section names and to add this feature to sections when EMBER data are transformed to the ontology. The list of usual section names is available in our GitHub repository as \texttt{section\_names.json}.



\section{PE Malware Ontology}
\label{sec:ontology}

\begin{figure}
    \centering
    \includegraphics[width=\textwidth]{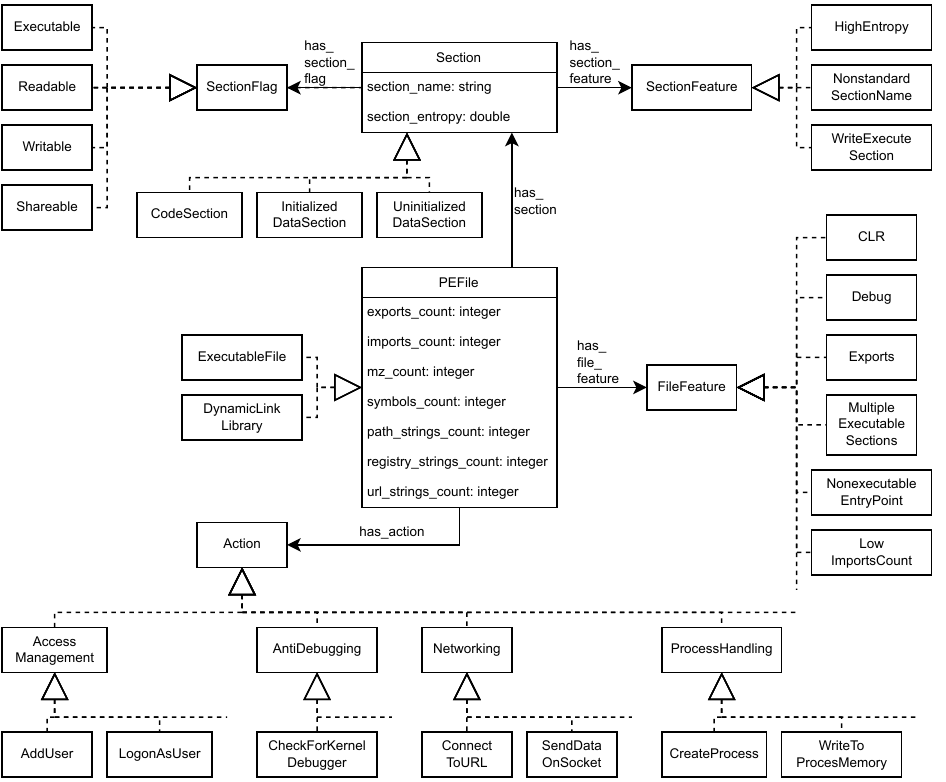}
    \caption{Core classes of the PE Malware Ontology and their properties}
    \label{fig:ontology}
\end{figure}


We now introduce the PE Malware Ontology, a formal conceptualization of the domain of static PE malware analysis in the OWL~2 ontology language, briefly introduced in Section~\ref{sec:prelim}. While we used the structure of \ac{EMBER} (see Section~\ref{sec:dataset}) as the starting point, it should be noted that our ontology is not a direct mapping of this or any other individual dataset.  
Since one of the most important goals was interpretability, we included only those properties and features that make sense from a malware expert's point of view
\cite{sikorski2012practical, kleymenov2022mastering}.

For this reason, we did not use various features such as file sizes, linker versions, byte histograms, etc. In theory, there may be patterns in the data involving such features and various classifiers can learn them, but from the expert's perspective, such features would hardly provide meaningful, easily interpretable explanations as to why a particular sample is malware or benign. Also, the use of such features could subsequently cause the trained classifiers to be prone to rather trivial adversarial attacks~\citep{suciu2019exploring}. 

Our ontology is partially depicted in Figure~\ref{fig:ontology}. The core classes and properties are fully depicted, but only an illustrative sample of file features and actions is shown for brevity. The ontology contains 195~classes, 6~object properties, and 9~data properties in total. The ontology is available in the \texttt{pe\_malware\_ontology.owl} file in our GitHub repository (see Section~\ref{sec:introduction}).

\subsection{PE files}
\label{sec:pefiles}

Each sample of a Portable Executable (PE) file \citep{bridge2022pe} from the original dataset is represented by an instance of the \texttt{PEFile} class, central to our ontology. Each such instance is further classified either as a proper executable (EXE) in the \texttt{ExecutableFile} subclass of \texttt{PEFile} or as a shared library (DLL) in the \texttt{DynamicLinkLibrary} subclass. The respective subclass is determined from the sample's COFF header available in the \ac{EMBER} dataset (see Listing~\ref{lst:ember}).

PE files structurally consist of sections, possess features relevant to malware detection, and may perform actions (based on their imported functions) when executed. These entities are represented by instances of the respective classes in the ontology (\verb+Section+, \verb+FileFeature+, \verb+Action+) which are linked to a \texttt{PEFile} instance via the \verb+has_section+, \verb+has_file_feature+, and \verb+has_action+ object property, respectively. These object properties and their ranges are summarized in Table~\ref{tab:PEFile}.

\begin{table}[bthp]\small
	\caption{\texttt{PEFile} properties}\label{tab:PEFile}
	\centering
    \setlength{\tabcolsep}{.85\tabcolsep}
    \begin{tabular}[t]{ >{\ttfamily} l >{\ttfamily} l @{\hspace{2\tabcolsep}} }
    \toprule
    \bf Object property & \bf Range
    \\
    \midrule
    has\_action & Action
    \\
    has\_file\_feature & FileFeature
    \\
    has\_section & Section
    \\
    \\
    \\
    \\
    \\
    \bottomrule
    \end{tabular}%
    \begin{tabular}[t]{ >{\tt} l >{\tt} l @{\hspace{2\arraycolsep}} }
    \toprule
    \bf Data property & \bf Range
    \\
    \midrule
    exports\_count & xsd:integer
    \\
    imports\_count & xsd:integer
    \\
    mz\_count & xsd:integer
    \\
    path\_strings\_count & xsd:integer
    \\
    symbols\_count & xsd:integer
    \\
    registry\_strings\_count & xsd:integer
    \\
    url\_strings\_count & xsd:integer
    \\
    \bottomrule
    \end{tabular}%
\end{table}

The \texttt{PEFile} class is also the domain of 7~data properties (see Table~\ref{tab:PEFile} again) which provide information on the numbers of imported and exported functions, debugging symbols, and MZ headers (mapped from the \texttt{general} property of JSON descriptions of samples), as well as the numbers of path, registry, and URL strings (mapped from the \texttt{strings} property). Similarly to features (cf.\ Section~\ref{sec:computed-features}), data properties were selected to be both easily interpretable and relevant to malware detection. For instance, malicious files usually expose no debugging symbols, packed malware tends to imports very few API functions and may contain multiple MZ headers. Malware may also use a number of URLs (for, e.g., C\&C communication, downloading additional payloads, etc.) or registry keys (for, e.g., achieving persistence, privilege escalation, etc.).

\subsection{File features}
\label{sec:filefeatures}

\begin{table}[htbp]\footnotesize
	\caption{PE file features}
	\centering
    \setlength{\tabcolsep}{.5em}
	\begin{tabularx}{\textwidth}{l >{\RaggedRight\hangindent=1em} X}
		\toprule
		\textbf{File feature class}
        & \textbf{Meaning (corresponding JSON sample property or~equivalent OWL~2 description)} \\
		\midrule
        \addlinespace
        \multicolumn{2}{l}{\textbf{Directly represented features}}
        \\\addlinespace[.5\defaultaddspace]
		\texttt{Debug}
        & Contains a debug section
            (\texttt{general.has\_debug}) \\
		\texttt{Relocations}
        & Contains a relocation section
            (\texttt{general.has\_relocations}) \\
		\texttt{Resources}
        & Contains resources (fonts, images, etc.;\NL
            \texttt{general.has\_resources}) \\
	    \texttt{Signature}
        & Is digitally signed
            (\texttt{general.has\_signature}) \\
		\texttt{TLS}
        & Includes a Thread Location Storage section
            (possibly a secret entry point;
            \texttt{general.has\_tls})
        \\\addlinespace
		\multicolumn{2}{l}{\textbf{Pre-processed features}}
        \\\addlinespace[.5\defaultaddspace]
        \texttt{CLR}
        & Contains a Common Language Runtime data directory
            (used in .NET binaries) \\
		\texttt{NonexecutableEntryPoint}
        & Entry point is not in an executable section
        \\\addlinespace
		\multicolumn{2}{l}{\textbf{Derived features}}
        \\\addlinespace[.5\defaultaddspace]
	    \texttt{Exports}
        & Exports some functions (mostly in DLLs;\NL
            $\exists\texttt{exports\_count}.\,\texttt{xsd:integer}[> 0]$) \\
		\makecell[tl]{\texttt{MultipleExecutable}\\\texttt{Sections}}
        & Has multiple sections with the executable flag
            (${\geqslant}2\,\texttt{has\_section}.\,
              \exists\texttt{has_section\_flag}.\,\texttt{Executable})$) \\
		\texttt{LowImportsCount}
        & The number of imported functions is smaller than the threshold value
            ($\exists\texttt{imports\_count}.\,\NL
              \texttt{xsd:integer}[< \texttt{\textit{imports\_threshold}}]$) \\
		\texttt{NonstandardMZ}
        & Has no or more than one MZ headers
            (possibly contains an embedded PE file;
            $\exists\texttt{mz\_count}.\,\texttt{xsd:integer}[> 1]$) \\
		\texttt{PathStrings}
        & Contains strings defining paths\NL
            ($\exists\texttt{path\_strings\_count}.\,
                \texttt{xsd:integer}[> 0]$) \\
		\texttt{RegistryStrings}
        & Contains strings defining registry keys
            ($\exists\texttt{registry\_strings\_count}.\,
                \texttt{xsd:integer}[> 0]$) \\
		\texttt{Symbols}
        & Has COFF debug symbols
            (deprecated in executables;
            $\exists\texttt{symbols\_count}.\,
                \texttt{xsd:integer}[> 0]$) \\
	    \texttt{URLStrings}
        & Contains strings defining URLs
            ($\exists\texttt{url\_strings\_count}.\,
                \texttt{xsd:integer}[> 0]$) \\
		\bottomrule
	\end{tabularx}
	\label{tab:filefeatures}
\end{table}

The \texttt{FileFeature} class has 15~subclasses representing various qualitative features of PE files that are, based on expert knowledge \cite{sikorski2012practical, kleymenov2022mastering}, relevant to malware detection. They are enumerated in Table~\ref{tab:filefeatures}.
These features are sorted into the three categories detailed in Section~\ref{sec:computed-features} and verbally described.
Moreover, for each directly represented feature, the table contains the respective JSON sample property from the EMBER dataset.
For each derived feature, the table contains a logical expression, in the DL syntax, deriving it from the PE file properties.
The pre-processed features were described in detail in Section~\ref{sec:computed-features}. 

Each subclass of \texttt{FileFeature} has a single prototypical instance. All \verb+PEFile+ instances that possess a feature are connected via the \texttt{has\_file\_\allowbreak feature} object property to the feature's prototypical instance.
Although this way of modeling features may be uncommon, it allows for easily readable descriptions of possible malware, e.g., $\texttt{ExecutableFile} \sqcap \exists\texttt{has\_file\_feature}.\allowbreak\{ \texttt{multiple\_executable\_sections} \}$ -- to be read as ``an executable file with the feature of having multiple executable sections.''

\subsection{Sections}
\label{sec:sections}

The instances of the \texttt{Section} class represent the sections of PE file samples. A \texttt{PEFile} instance is linked to all its sections via the \texttt{has\_section} property. Sections are further classified based on the type of data found in the given section into one of \verb+Section+'s three subclasses: \texttt{CodeSection}, \texttt{InitializedDataSection}, and \texttt{UninitializedDataSection}.

\texttt{Section} instances are linked to their permission flags and section features by the respective object properties shown in Table~\ref{tab:Section}. Furthermore, \texttt{Section} instances have data properties assigning them their names and the entropies of their content. The significance of this data for malware detection has been discussed in Section~\ref{sec:computed-features}.

\begin{table}[htbp]\small
	\caption{\texttt{Section} properties}\label{tab:Section}
	\centering
    \begin{tabular}[t]{ >{\tt} l >{\tt} l @{\hspace{2\arraycolsep}} }
    \toprule
    \bf Object property & \bf Range
    \\
    \midrule
    has\_section\_feature & SectionFeature
    \\
    has\_section\_flag & SectionFlag
    \\
    \bottomrule
    \end{tabular}%
    \begin{tabular}[t]{ @{\hspace{2\arraycolsep}} >{\tt} l >{\tt} l }
    \toprule
    \bf Data property & \bf Range
    \\
    \midrule
    section\_entropy & xsd:double
    \\
    section\_name & xsd:string
    \\
    \bottomrule
    \end{tabular}
\end{table}

\subsection{Section flags}
\label{sec:sectionflags}

The \texttt{SectionFlag} class classifies the flags given to PE file sections. In the \ac{EMBER} dataset, these flags are included in the \verb+props+ property of each section. From the malware-detection point of view, the most interesting of them are the flags controlling how processes executing the PE file may access (read, write, execute) and share the memory region into which this section is mapped. These flags can be, in theory, combined arbitrarily. Only these four flags are represented in the ontology, by the \verb+Executable+, \verb+Readable+, \verb+Writable+, and \verb+Shareable+ subclasses of \texttt{SectionFlag}. Similarly to \texttt{FileFeature}'s subclasses, each subclass of \texttt{SectionFlag} has a prototypical instance to which the sections with the respective flag are linked via the \verb+has_section_flag+ object property.

\subsection{Section features}
\label{sec:sectionfeatures}

\verb+SectionFeature+'s subclasses represent the features of sections that, again, based on expert knowledge \cite{sikorski2012practical, kleymenov2022mastering}, are relevant for malware detection. They are listed in Table~\ref{tab:sectionfeatures}. All of these features are derived, and their values are computed from the section's data properties and flags, as detailed in Section~\ref{sec:computed-features}.

The instances of \texttt{Section} are connected to the prototypical instances of the subclasses of \texttt{SectionFeature} using the \texttt{has\_section\_feature} property.

\begin{table}[htbp]\small
	\caption{Section features}
	\centering
	\begin{tabularx}{\textwidth}{l >{\RaggedRight\hangindent=1em} X}
		\toprule
		\textbf{Section feature class}
        & \textbf{Meaning (equivalent OWL~2 description)} \\
		\midrule
		\texttt{HighEntropy} &
            The value of section's entropy is larger than the threshold value
            ($\exists\texttt{section\_entropy}.\,\allowbreak
                \texttt{xsd:double}[> \textit{\ttfamily entropy\_threshold\/}]$)   \\
		\texttt{NonstandardSectionName}
        & The section's name is not in the list of standard section names
            ($\exists\texttt{section\_name}.\,\allowbreak\lnot\{~\texttt{".text"}, \texttt{".data"}, \texttt{".rsrc"}, \ldots\ \}$)\\
	    \texttt{WriteExecuteSection}
        & Section has write and execute permissions
            ($\exists\texttt{has\_section\_flag}.\,\texttt{Writable}
                \sqcap \NL \phantom{(}
                \exists\texttt{has\_section\_flag}.\,\texttt{Executable}$)\\
		\bottomrule
	\end{tabularx}
	\label{tab:sectionfeatures}
\end{table}

\subsection{Actions}
\label{sec:actions}

\begin{table}[htbp]\small
	\caption{Intermediate action classes}
	\centering
    \setlength{\tabcolsep}{.5em}
	\begin{tabularx}{\textwidth}{l >{\RaggedRight\hangindent=1em} X}
		\toprule
		\rmfamily\textbf{Action class}    & \textbf{Meaning}  \\
		\midrule
		\texttt{AccessManagement} & Managing users on the system (adding new user, enumerating existing users, etc.)    \\
		\texttt{AntiDebugging}     & Debugger detection techniques  \\
	    \texttt{Cryptography}    &  Encrypting/decrypting files, generating keys, etc.      \\
	    \texttt{DirectoryHandling}    & Manipulating with directories (creation, deletion, etc.)       \\
	    \texttt{DiskManagement}    & Mounting/unmounting disks, enumerating existing disks       \\
	    \texttt{FileHandling}    & Manipulating with files (creation, deletion, etc.)       \\
	    \makecell[tl]{\texttt{InterProcess}\\[-0.2em]\texttt{Communication}}    & Communication between processes (named pipes)       \\
	    \texttt{LibraryHandling}    & Loading library into running processes       \\
	    \texttt{Networking}    & Various networking activities (connecting to a socket, sending DNS requests)       \\
	    \texttt{ProcessHandling}    &  Various APIs for process handling, including creation or modifying process memory      \\
	    \texttt{RegistryHandling}    & Enumerating registry keys, writing new values, etc.       \\
	    \texttt{ResourceSharing}    & Manipulating with resources shared over network       \\
	    \texttt{ServiceHandling}    & Manipulating with systems's services       \\
	    \makecell[tl]{\texttt{Synchronization}\\[-0.2em]\texttt{PrimitivesHandling}}    & Handling mutexes/semaphores       \\
	    \texttt{SystemManipulation}    & Various APIs for obtaining system information       \\
		\texttt{ThreadHandling}    & Creating remote threads, enumerating running threads       \\
		\texttt{WindowHandling}    & Window manipulation APIs (creating new windows, dialog boxes, etc.)       \\
		\bottomrule
	\end{tabularx}
	\label{tab:actions}
\end{table}

Instances of the class \texttt{Action} represent actions that may be taken by a process executing the code from a PE file. Actions are classified in 139~leaf%
\footnote{A \emph{leaf} class is one with no named subclasses except itself.}
subclasses of \texttt{Action} that are aligned with \ac{MAEC}'s Malware actions vocabulary as described in Section~\ref{sec:standardization}. In order to aid generalization during classifier learning, we have added 17~categories of actions. Each category is represented by an intermediate class (e.g., \texttt{Process\allowbreak Handling}; see Figure~\ref{fig:ontology})~-- a direct subclass of \texttt{Action} and a direct superclass of the leaf \ac{MAEC}-based action classes falling into the category (e.g., \texttt{Create\allowbreak Process}, \texttt{Write\allowbreak To\allowbreak Process\allowbreak Memory}). All 17~intermediate classes together with the description of the category they represent are listed in Table~\ref{tab:actions}. 

In the current version of our ontology, each leaf action class has a prototypical instance and all \verb+PEFile+ instances that may perform this action are connected to it by the \texttt{has\_\allowbreak action} object property. This connection is created by mapping imported functions from the \verb+imports+ property of samples in the \ac{EMBER} dataset to the corresponding \ac{MAEC} malware actions if possible. For example, a \texttt{PEFile} instance representing a sample importing an API function that creates a process is connected to a prototypical instance \texttt{create-process} of the \texttt{Create\allowbreak Process} class via the \texttt{has\_\allowbreak action} property. Of course, the fact that a sample imports an API function does not necessarily mean that the function will be called when this sample is run. This, however, is a general limitation of static malware analysis and not of the designed ontology. Dynamic malware analysis would have to be used to make the data more precise.

Alternatively, if data from dynamic malware analysis is available, typically obtained by running the samples in a sandbox environment, actions actually performed by a sample could be represented by non-prototypical instances of \texttt{Action}, classified in the appropriate leaf subclass and further characterized by additional properties, such as the parameters, the previously and the subsequently taken action, or the timestamp. Extending our ontology to cover such dynamic data may be interesting for future work. 


\subsection{Logical properties and modules}

The PE Malware Ontology is relatively lightweight in terms of expressivity and reasoning complexity. It can be expressed in the description logic DL-Lite\textsubscript{core}($D$), the basis of the OWL\,2~QL profile~\citep{motik2012owl2profiles}. This profile is aimed at efficient query answering with respect to the size of the data (ABox, see Section~\ref{sec:prelim}). Our ontology's main purpose is to provide a shared vocabulary for describing the static features of PE malware and benign samples without adding much background knowledge.
Perhaps the only interesting derivations the basic ontology enables are those of more abstract action classes from the specific ones assigned to samples via the \texttt{has\_action} property. Being lightweight, the ontology facilitates efficient retrieval of samples for processing in ML applications.

We have created a module adding disjointness axioms to the basic ontology (\texttt{disjointness.owl}). The module remains in the OWL\,2~QL profile. With these axioms, it is also possible to derive further disjointness consequences. For instance, since \texttt{Section} and \texttt{Action} are disjoint, so are all pairs of their respective subclasses such as \texttt{CodeSection} and \texttt{Networking}. This enables SML tools to reduce the search space since they can infer that descriptions such as $\texttt{ExecutableFile} \sqcap \exists\texttt{has\_action}.\texttt{CodeSection}$ (an EXE file that may perform an action categorized as code \emph{section}) match no samples.

Additionally, we offer a module consisting of logical axioms defining the derived file and section features (\texttt{feature-definitions.owl}), using the DL expressions seen in Tables \ref{tab:filefeatures} and~\ref{tab:sectionfeatures}. These axioms require the full expressive power of OWL\,2~DL and the reasoning with them is thus more complex.
The definitions may have an effect in SML applications, since a tool capable of OWL\,2~DL reasoning may, hypothetically, reduce the search space by skipping expressions equivalent to a feature definition once they have considered the respective feature. The real-world impact of including the definitions requires further investigation.

The datasets that we provide (see Section~\ref{sec:fractions}) already contain assertions assigning the derived features to samples, which was done specifically to help avoid the reasoning costs and facilitate the use of tools that lack the required reasoning capabilities. In use cases aiming at fully exploiting the feature definitions, these assertions thus need to be filtered out from the datasets. Such a filtering is easily implemented since in the basic ontology, the derived features are annotated by the property \texttt{derived\_as} with the textual form of the defining expressions.

\begin{lstlisting}[language=turtle, caption={PE file sample description in the PE Malware Ontology.}, label={lst:pemalware:turtle}, float=bt]
@prefix : <https://orbis-security.com/pe-malware-ontology#> .
:nonstandard_mz rdf:type :NonstandardMZ , owl:NamedIndividual .
:create-window  rdf:type :CreateWindow , owl:NamedIndividual .
...

:aba129a3d1ba9d307dad05617f66d8e7
    rdf:type :ExecutableFile , owl:NamedIndividual ;
    :mz_count 11 ;
    :exports_count 0 ;      :imports_count 17 ;         :symbols_count 0 ;
    :path_strings_count 0 ; :registry_strings_count 0 ; :url_strings_count 9 ;
    :has_section
        :code_aba129a3d1ba9d307dad05617f66d8e7 ,
        :data_aba129a3d1ba9d307dad05617f66d8e7 ,
        :idata_aba129a3d1ba9d307dad05617f66d8e7 ,
        :rdata_aba129a3d1ba9d307dad05617f66d8e7 ,
        :reloc_aba129a3d1ba9d307dad05617f66d8e7 ,
        :rsrc_aba129a3d1ba9d307dad05617f66d8e7 ;
    :has_file_feature
        :nonstandard_mz ,   :relocations ,  :resources ,
        :tls ,              :url_strings ;
    :has_action
        :create-window ,            :delete-critical-section ,
        :read-registry-key-value ,  :release-critical-section ,
        :sleep-process .

:code_aba129a3d1ba9d307dad05617f66d8e7
    rdf:type :CodeSection , owl:NamedIndividual ;
    :section_name "code" ;
    :section_entropy "6.532932639432919"^^xsd:double ;
    :has_section_flag :executable , :readable .

:reloc_aba129a3d1ba9d307dad05617f66d8e7
    rdf:type :InitializedDataSection , owl:NamedIndividual ;
    :section_name "reloc" .
    :section_entropy "6.672072010670469"^^xsd:double ;
    :has_section_flag :readable , :shareable .
...
\end{lstlisting}

\subsection{Example of a semantic PE file sample description}

Listing~\ref{lst:pemalware:turtle} shows a part of the semantic description of the sample in Listing~\ref{lst:ember} within the PE Malware Ontology in the Turtle syntax~\citep{becket2014turtle}.
Individuals, classes, and properties are all referred to by IRIs (Internationalized Resource Identifiers). The individual representing the sample is referred to by an IRI derived from its MD5 sum (\texttt{:aba...8e7} in this case). IRIs referring to its sections are composed from section names and the sample's MD5 sum (e.g., \texttt{:code\_aba...8e7}). The first line of Listing~\ref{lst:pemalware:turtle} declares the common prefix of these IRIs.

\looseness=-1
Recall from Section~\ref{sec:prelim} that data in ontologies are given in the form of class and property assertions. In the Turtle syntax, a class assertion $C(a)$ is written as the triple ``\mbox{\texttt{a rdf:type C.}}'' and a property assertion $P(a,b)$ is written as the triple ``\mbox{\texttt{a P b.}}''. Multiple triples describing a single entity can be combined for brevity. For instance, the assertions $C(a)$, $D(a)$, $P(a,b)$, $P(a,c)$, $R(a,d)$ can be written as ``\mbox{\texttt{a rdf:type C, D; P b, c; R d.}}''.

The second and third line of Listing~\ref{lst:pemalware:turtle} describe examples of the prototypical instances (\texttt{non\-stand\-ard\_\allowbreak mz} and \texttt{create-window}) of the file feature and action classes (\texttt{NonstandardMZ} and \texttt{CreateWindow}, respectively).

The PE file sample \texttt{:aba...8e7} is then described by asserting its class, its data properties, and its object properties (cf.\ Section~\ref{sec:pefiles}). The latter connect the PE file sample to individuals representing its sections (\texttt{has\_section}), to the prototypical instances of its features (\texttt{has\_file\_feature}) and actions it may perform (\texttt{has\_actions}).

Finally, descriptions of two examples of this sample's sections are shown: the code and the resources sections. Neither of these sections has any distinguished section features, since their names are standard and their respective contents entropies are below the high entropy threshold.

\section{Datasets}
\label{sec:fractions}

One of the important goals was to create datasets that would allow unambiguous reproduction of experimental results. For this reason, we generated multiple datasets of different sizes. Each dataset can be downloaded and exactly referred to when used in experiments. This way, even independent experiments may be more directly compared. Altogether, we publish 31~datasets exported from the EMBER data and 5~datasets exported from the \ac{SOREL} data.

\subsection{EMBER datasets}
\label{sec:onto-datasets}

Each dataset contains 50~\% positive samples (malware) and 50~\% negative samples (benign). We generated datasets of various sizes from 1000 to 800,000 (i.e., the entire \ac{EMBER} dataset excluding unlabeled samples). The main reason was that concept-learning algorithms are computationally more demanding than some traditional ML algorithms, and for that reason, it can be useful to have less robust datasets available. We also generated several variants for each fractional dataset size, which were randomly assembled from all \ac{EMBER} data. More details can be seen in Table~\ref{tab:our-ontologies}.

\begin{table*}[bt]
    \caption{Approximate metrics of the PE malware ontological datasets \texttt{dataset\_\textit{N}\_\textit{size}.owl}.}
    \label{tab:our-ontologies}
    \scriptsize\centering\setlength{\tabcolsep}{0.4em}
    \begin{tabular}{ r *{11}{r} }
    \toprule
    \multirow[b]{2}{*}{\makecell[cc]{Dataset\\\texttt{\textit{size}}}}
    &\multicolumn{3}{c}{Examples}
    &
    &\multicolumn{2}{c}{Properties}
    &
    &
    &\multicolumn{3}{c}{Assertions}
    \\
    \cmidrule(lr){2-4}
    \cmidrule(lr){6-7}
    \cmidrule(lr){10-12}
    & 
    \multicolumn{1}{c}{All} &
    \multicolumn{1}{c}{Pos.} &
    \multicolumn{1}{c}{Neg.} &
    \multicolumn{1}{c}{Classes} &
    \multicolumn{1}{c}{Obj.}& 
    \multicolumn{1}{c}{Data}& 
    \multicolumn{1}{c}{Indivs.} & 
    \multicolumn{1}{c}{Axioms} &
    \multicolumn{1}{c}{Class} &
    \multicolumn{1}{c}{Obj.\,prop.} &
    \multicolumn{1}{c}{Data prop.}
    \\
    \midrule
    \multicolumn{12}{l}{EMBER-based datasets}\\
    1000 &
    1\,k & 500 & 500 & 195 & 6 & 10 & 5.8\,k & 63\,k 
    & 5.8\,k & 34\,k & 16\,k
    \\
    10000 &
    10\,k & 5\,k & 5\,k  & 195 & 6 & 10 & 57\,k & 621\,k 
    & 57\,k & 343\,k & 164\,k
    \\
    100000 &
    100\,k & 50\,k & 50\,k & 195 & 6 & 10 & 567\,k & 6,189\,k
    & 568\,k & 3,418\,k & 1,634\,k
    \\
    800000 &
    800\,k & 400\,k & 400\,k & 195 & 6 & 10 & 4,537\,k & 49,506\,k
    & 4,545\,k & 27,348\,k & 13,075\,k
    \\
    \addlinespace
    \multicolumn{12}{l}{SoReL-based datasets}\\
    1000 &
    1\,k & 500 & 500 & 195 & 6 & 10 & 6.1\,k & 50\,k %
    & 6.3\,k & 24\,k & 13\,k
    \\
    \bottomrule
    \end{tabular}
\end{table*}

There are three files for each dataset (where \texttt{N} denotes variant number and \texttt{M} denotes dataset size):

\begin{description}
    \item[\texttt{dataset\_N\_M.owl}:] the ontology filled with OWL individuals which contains \texttt{M} malware/benign samples;	
    \item[\texttt{dataset\_N\_M\_raw.json}:] the original JSON samples from the \ac{EMBER} dataset (including the unused features) from which the ontological dataset was generated;
    \item[\texttt{dataset\_N\_M\_examples.json}:] the list of the positive and negative samples in the given dataset in a JSON format.
\end{description}

We refrain from publishing a predefined split of the datasets into the training and test parts, as committing to one split may lead to a lucky or unlucky setting for certain learning algorithms or even to overfitting the given predefined split. Instead, we propose to follow the $k$-fold cross-evaluation methodology \citep{kfold} in which case no predefined split is needed (cf.\ Section~\ref{sec:case_study:methodology}).

\subsection{\ac{SOREL} Datasets}
As a demonstration of the broader applicability of our PE Malware Ontology, we also used this ontology to prepare 5 small, mutually disjoint datasets based on fragments of \ac{SOREL}. Each of these datasets comprises 1000 samples randomly selected from the entire \ac{SOREL} in a way that the resulting dataset consists of $250+250$ malicious EXE${}+{}$DLL samples and $250+250$ benign EXE${}+{}$DLL samples.
For each dataset, we again provide three files: \texttt{dataset\_N\_1000.\allowbreak owl}, \texttt{dataset\_N\_1000\_\allowbreak raw.\allowbreak json}, and \texttt{dataset\_N\_1000\_\allowbreak examples.\allowbreak owl}, with the same type of contents as in the case of our EMBER datasets discussed previously.

We decided to ensure the above-described balance between the numbers of EXEs and and DLLs, as well as malicious and benign samples among both file categories because we noticed that the ratio of EXEs to DLLs in \ac{SOREL} is approximately 4 to 1 and that barely 10~\% of DLL samples in \ac{SOREL} are malicious. Keeping such proportions would render it impossible to capture enough representative knowledge about DLLs (especially malicious ones) in our tiny datasets and thus introduce a bias towards focusing on features of malicious EXEs.

We also have to mention that \ac{SOREL} does not store the results from a string analysis of the PE files, since none was performed by the authors. Therefore, the statistics for MZ strings, URL-path strings, etc.\ are not included in the created semantic datasets either. In addition, we have observed that some PE samples in \ac{SOREL} include sections with multiple content types assigned. We categorized each such section into all the respective subclasses of \texttt{Section}. As a result, the average number of class assertions in the ontological datasets is slightly higher than that of the individuals (see Table~\ref{tab:our-ontologies}).

\section{Case Study}
\label{sec:case-study}

We will now present a preliminary case study with the aim to underpin the approach and demonstrate its general viability.
We first show baseline results obtained using LIME and SHAP. Then to demonstrate the potential of SML methods, we apply concept learning \cite{DL-Learner}.
The published datasets could potentially be exploited by diverse SML methods \cite{westphal2019sml}. 

\subsection{Baseline}

LIME \cite{LIME} is a model-agnostic explanation method that focuses on generating local explanations for individual predictions. LIME was applied on an Mult-Layer Perceptron (MLP)-based malware classifier binarized version of our EMBER ontology as detailed by Moj\v{z}i\v{s} and Kenyeres \cite{MojisKenyeres2023}. The model has two ReLU-activated dense layers (512 and 256 units) with dropout (0.4 and 0.2) for regularization, ending in a sigmoid layer for malware probability prediction. It was trained using the Adam optimizer (lr=0.001) with binary cross-entropy loss. Early stopping (patience=3) and a fixed seed (42) were used to ensure reproducibility.

\begin{figure}[h!]
    \centering
    \includegraphics[width=\linewidth]{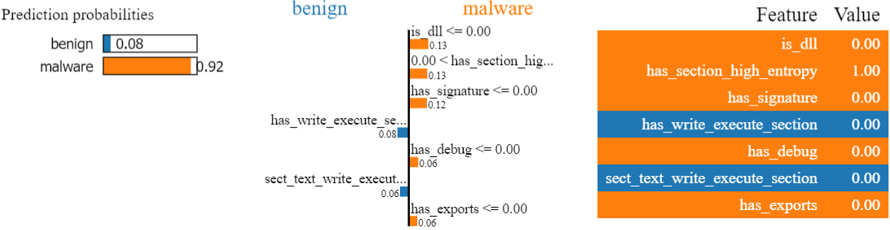} 
    \caption{LIME explanation for a single malware instance -- the bar chart shows the feature contributions toward the classification as \textit{Malware} (orange) or \textit{Benign} (blue)}
    \label{fig:lime_exp}
\end{figure}

LIME generates local explanations for each specific input instance. One such explanation is visualized in Figure~\ref{fig:lime_exp}. For this instance, the prediction probabilities were 0.08 for \textit{Benign} and 0.92 for \textit{Malware}. The explanation provides insights into the contributions of individual features to the model’s prediction. 
The key features influencing the prediction are visualized in Figure~\ref{fig:lime_exp}. Features contributing positively toward the classification as \textit{Malware} are shown in orange, while features supporting the \textit{Benign} classification are shown in blue. For this example, features such as \texttt{is\_dll} with value = \texttt{0} (indicating a non-dynamic library link), \texttt{has\_section\_high\_entropy} with value = \texttt{1} (indicating the file has at least one section with high entropy), and \texttt{has\_signature} with value = \texttt{0} (indicating the absence of a valid digital signature) were identified as the most important contributors towards the classification as malware. Although LIME explanations were insightful for individual instances, they do not provide a global understanding of the model’s decision-making process.

\begin{figure}[h!]
\centering
\begin{minipage}{0.45\textwidth}
    \centering
    \includegraphics[width=\textwidth]{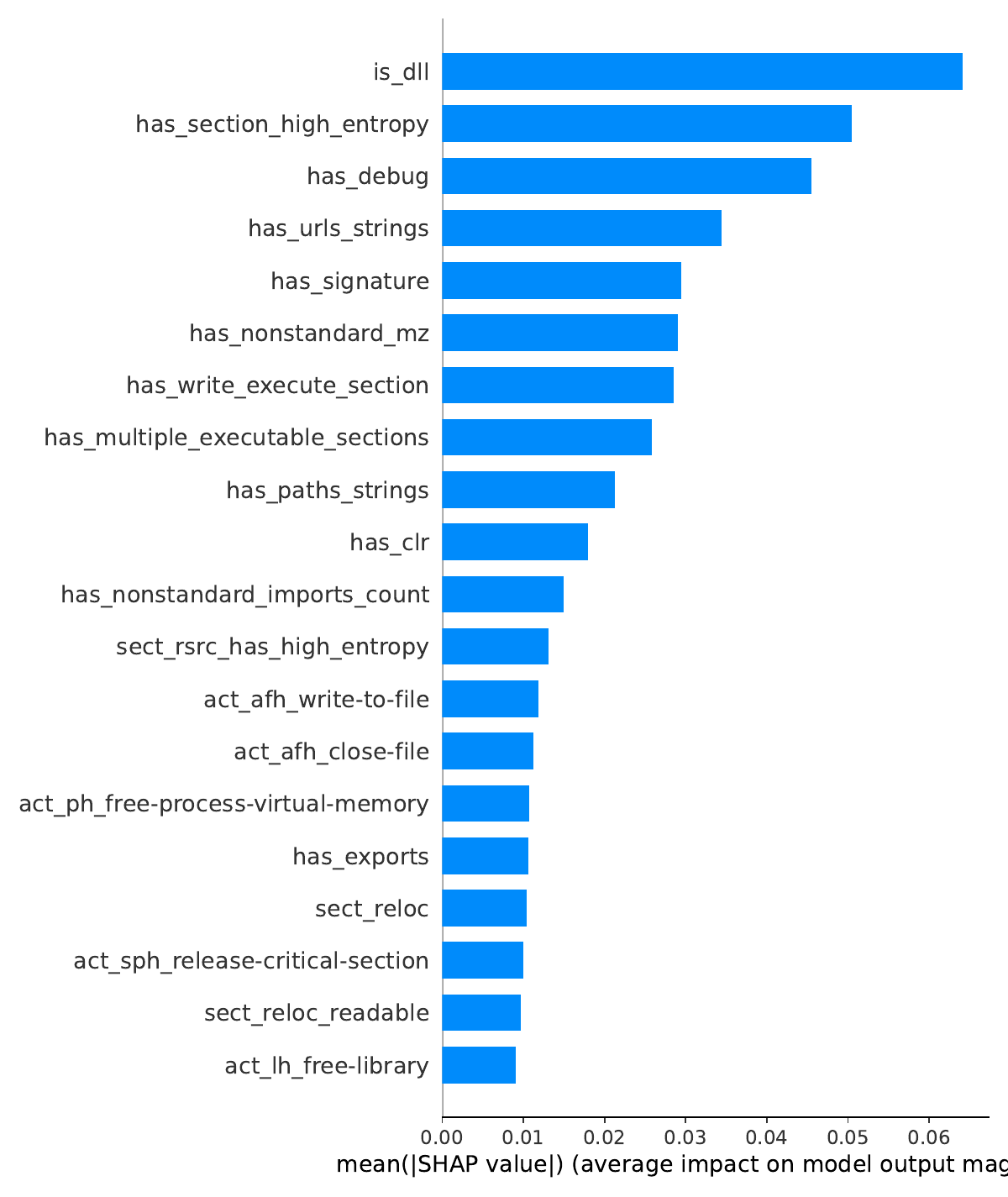}
    \caption{SHAP explanation (bar plot)}
    \label{fig:shap_exp}
\end{minipage}
\hfill
\begin{minipage}{0.45\textwidth}
    \centering
    \includegraphics[width=\textwidth]{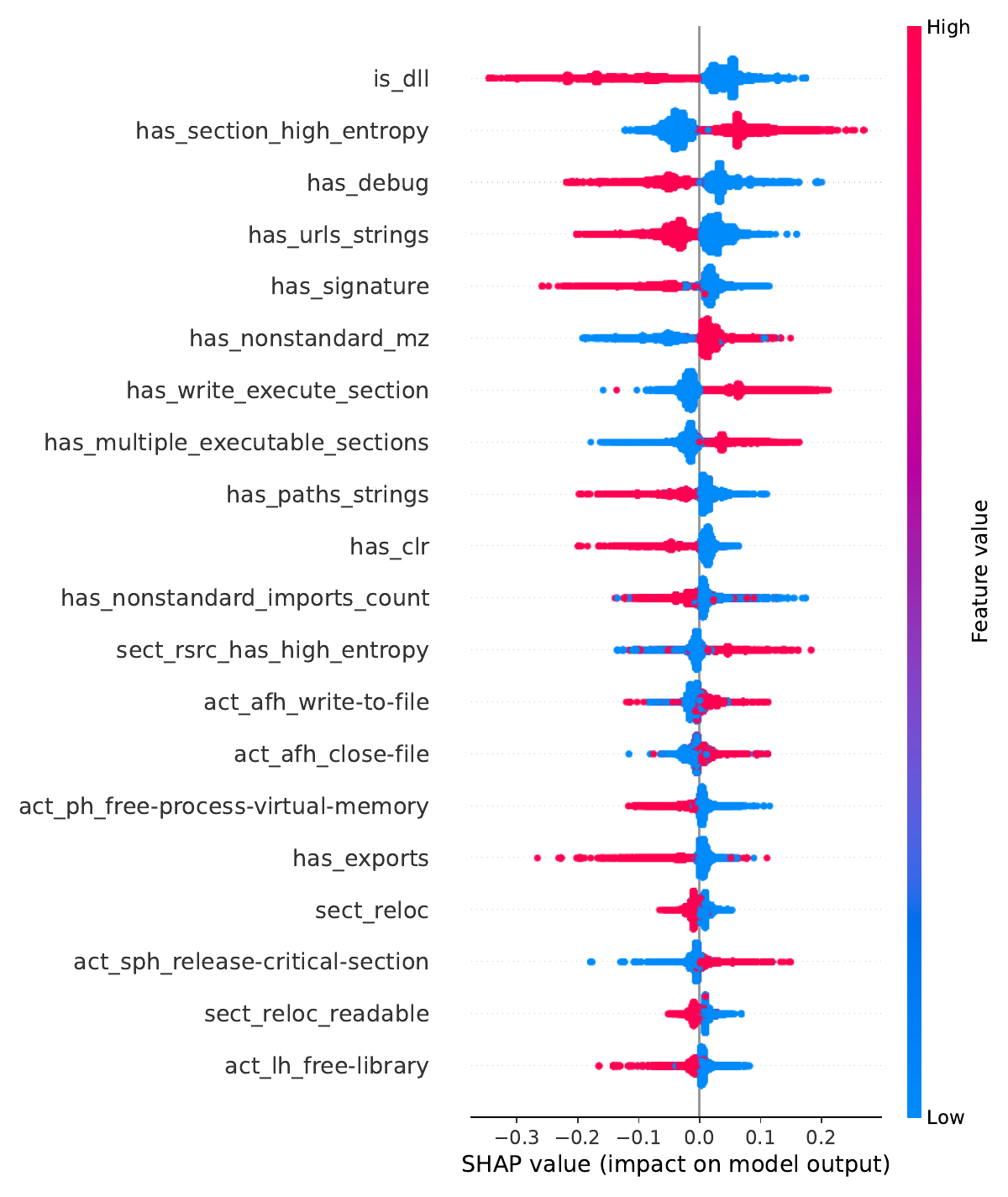}
    \caption{SHAP explanation (beeswarm plot)}
    \label{fig:shap_boxplot}
\end{minipage}
\end{figure}

In contrast, SHAP \cite{SHAP} is able to compute global explanations in terms of feature importance (so-called Shapley value) ranging over the whole dataset. Utilizing the same trained MLP model used for LIME, we applied SHAP to extract the global explanation using the test set. The results are depicted in Figure~\ref{fig:shap_exp}. The bar plot ranks features by their mean absolute SHAP values, with \texttt{is\_dll} being the most influential, followed by \texttt{has\_section\_high\_entropy}, \texttt{has\_debug}, and \texttt{has\_urls\_strings}. 

The box plot in Figure~\ref{fig:shap_boxplot} adds nuance by showing the distribution of SHAP values across all samples. Since the features are binary (0 or 1), red points represent a value of 1, while blue points represent 0. For \texttt{is\_dll}, a value of 1 is associated with a negative impact, as shown by the red points clustering on the left. In contrast, \texttt{is\_dll=0} has a neutral or slightly positive effect. Features like \texttt{has\_section\_high\_entropy} show a wide spread, influencing predictions both positively and negatively depending on the instance. This detailed analysis highlights the varying contributions of features across the dataset.

We point out that LIME and SHAP certainly offer relevant insights and enable the user to comprehend the relevance of each individual feature towards a given decision of the model (be it local or global) -- however the main drawback is that they are not able to capture deeper relations among the features and therefore the insight is limited. 

\subsection{Concept Learning}

Concept learning \cite{DL-Learner} applies ideas from  Inductive  Logic Programming (ILP) in the area of Description Logics (DL). 
Given an ontology $\mathcal{O}$, the task of concept learning \citep{lehmann2010concept} (also concept separability \citep{funk2019separability}) 
is to generate a complex concept description $C$ that characterizes the input sample $E = E^+ \cup E^-$ consisting of example instances $E^+$ and example non-instances $E^-$ of the target concept $C$ -- or, more formally -- such that $\mathcal{O} \models C(e)$ for all $e \in E^+$ and $\mathcal{O} \not\models C(e)$ for all $e \in E^-$. However, depending on the particular input sample~$E$, such a concept description may not exist or it may be too difficult to construct it, therefore practical concept-learning systems typically construct some approximation of $C$.

Generally, concept-learning algorithms search through the space of all possible concept descriptions, using (i) a \textit{refinement operator} \citep{lehmann2010learning}, which, given some input concept, returns a set of refined concepts to be searched next, and (ii) some \emph{heuristics} to control how the search space is traversed. Also, most typically the search starts from the $\top$ concept (the most general concept), and a downward refinement operator is used which always refines a given concept into a set of more specific concepts. Hence, the search space is organized into a tree.

We experimented with four learning algorithms: 
{OCEL}, {CELOE}, {PARCEL}, and {SPARCEL}, all implemented in \textit{DL-Learner} \citep{DL-Learner}, a state-of-the-art framework for supervised machine learning in description logics \citep{buhmann2018dl}. DL-Learner allowed us to target concepts up to $\mathcal{ALCOQ}$ DL expressivity in all four cases.

\textit{OCEL} (OWL Class Expression Learner)
is the most basic concept-learning algorithm. It uses the $\rho$ refinement operator \citep{lehmann2010learning}. The algorithm provides various techniques to cope with redundancy and infinity by revisiting nodes in the search tree several times and performing additional refinements.

\textit{CELOE} (Class Expression Learning for Ontology Engineering), a variant of OCEL, uses the same refinement operator but different heuristics, which are biased towards shorter concepts. Shorter concepts are often more useful for enhancing knowledge bases~\citep{lehmann2011celoe}, and they are also more easily understandable as explanations.

\textit{PARCEL} (Parallel Class Expression Learner)
uses a slightly different approach by finding partial definitions of the learning problem, which are then aggregated into a complete solution using disjunction \citep{tran2012approach}. Partial definition is class expression that covers at least one positive example. The main advantage of this approach is that it can be parallelized, where each working thread is refining different parts of the search tree.

\textit{SPARCEL} (Symmetric Parallel Class Expression Learner)
is a variation of the {PARCEL} algorithm \citep{tran2017parallel}. While \textit{PARCEL} focuses on finding partial definitions from the positive examples (descriptions that cover some of the positive examples and none of the negative examples), {SPARCEL} also tries to find partial definitions of the negative examples and combine them into the final solution (using the logical pattern ``$A$ and not $B$'').

Besides these algorithms, there are other approaches (not implemented in DL-Learner). \textsc{DL-Foil} \citep{fanizzi2018dlfoil} and its variation \textsc{DL-Focl} \citep{rizzo2020class} are FOIL-like (First-Order Inductive Learner) algorithms applied to description logics. 
\textsc{EvoLearner} \citep{Evolearner} uses an evolutionary approach to learn concepts. The \textsc{SPELL} system \citep{SPELL} relies on the PAC method and learns less expressive concepts in the tractable description logic $\mathcal{EL}$. The pFOIL-DL algorithm \citep{straccia2015pfoil}, on the other hand, extends to fuzzy description logics. \textsc{Fuzzy OWL-Boost} \citep{cardillo2022fuzzyOWLboost} and \textsc{Fuzzy PN-OWL} \cite{cardillo2023PN-OWL} adapt and extend the AdaBoost boosting algorithm to learn fuzzy concept inclusion axioms. 

\subsection{Methodology}
\label{sec:case_study:methodology}


Concept learning is computationally demanding; thus, in order to establish a baseline of results in this domain, we started with the 1\,k datasets from our suite (as characterized in Table~\ref{tab:our-ontologies}).
For the calibration of parameters, we randomly selected the dataset \texttt{dataset\_8\_1000.owl}.
For cross-validation, we selected the datasets 1--5 of the same size. 


We used the standard evaluation metrics \citep{fawcett2006introduction}
based on the total number~$P$ of malware (Positive) samples,
the total number~$N$ of benign (Negative) samples,
the numbers $\it TP$ and~$\it TN$ of samples
correctly classified as malware (True Positive) and benign (True Negative),
and the numbers $\it FP$ and~$\it FN$ of samples
incorrectly classified as malware (False Positive) and benign (False Negative),
respectively.
Obviously $P = \it TP + FN$ and $N = \it TN + FP$.
The metrics are then defined as usual:
\begin{align*}
    \it accuracy &= \frac{\it TP + TN}{P + N}
    &
    \it precision &= \it\frac{TP}{TP + FP}
	&
    \it recall &= \it\frac{TP}{TP + FN}
    \\
    \it FP\ rate &= \it\frac{FP}{FP + TN}
    &
	\it F1 &= \rlap{$\it\displaystyle\mathrm{2}\,\frac{precision \cdot recall}{precision + recall}$}
\end{align*}






We followed the $k$-fold cross-validation methodology \citep{kfold}: The dataset is split into $k$ equally sized subsets, where one subset is used for testing and the rest is used for training. The training process is repeated $k$ times, and the obtained results are averaged. This method is used for evaluating the generalization properties of a trained model and it also works quite well with smaller and limited datasets. Common values for $k$ are 3, 5, or 10. Due to computational complexity and the amount of experiments, we chose~$k=5$.

\begin{figure}[htbp]
\begin{subfigure}{.5\textwidth}
  \centering
  \includegraphics[width=.9\linewidth]{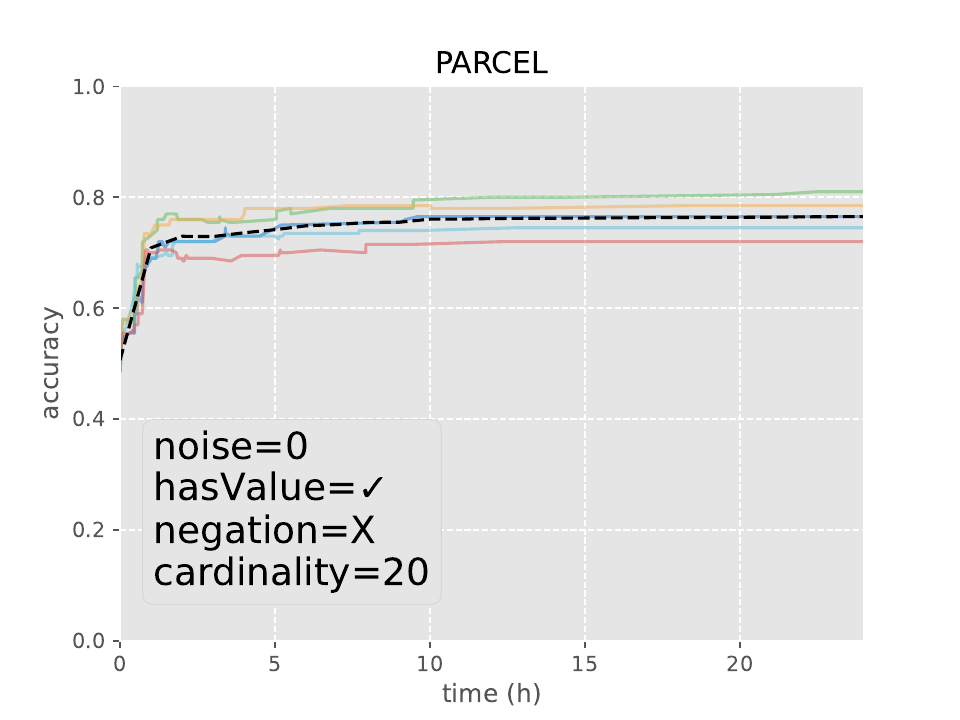}  
  \caption{}
\end{subfigure}\hfil
\begin{subfigure}{.5\textwidth}
  \centering
  \includegraphics[width=.9\linewidth]{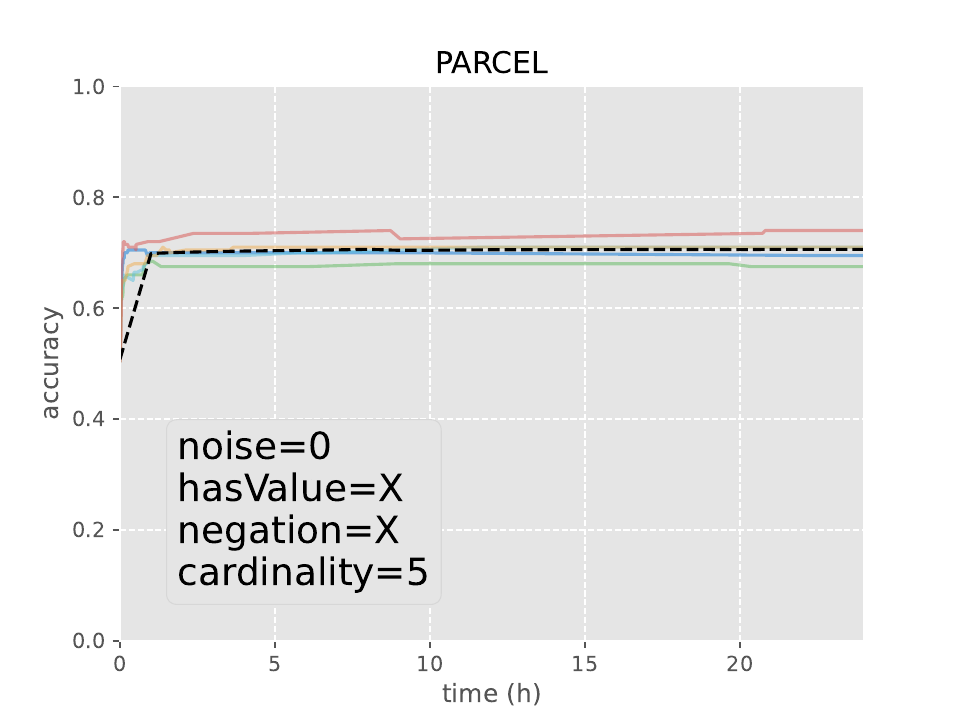}  
  \caption{}
\end{subfigure}

\leavevmode

\begin{subfigure}{.5\textwidth}
  \centering
  \includegraphics[width=.9\linewidth]{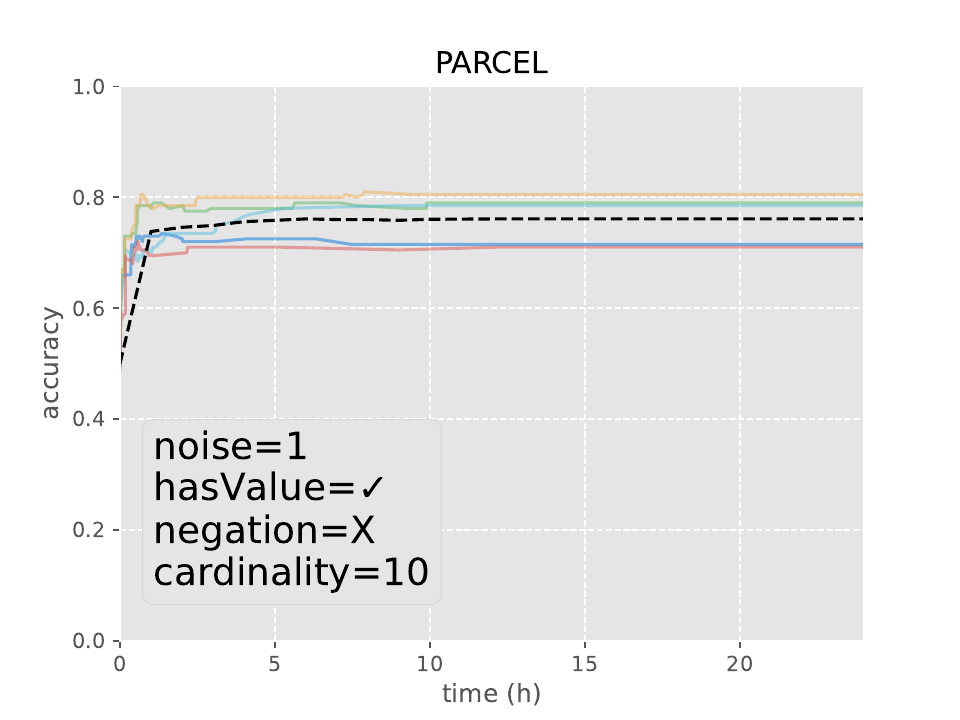}  
  \caption{}
\end{subfigure}\hfil
\begin{subfigure}{.5\textwidth}
  \centering
  \includegraphics[width=.9\linewidth]{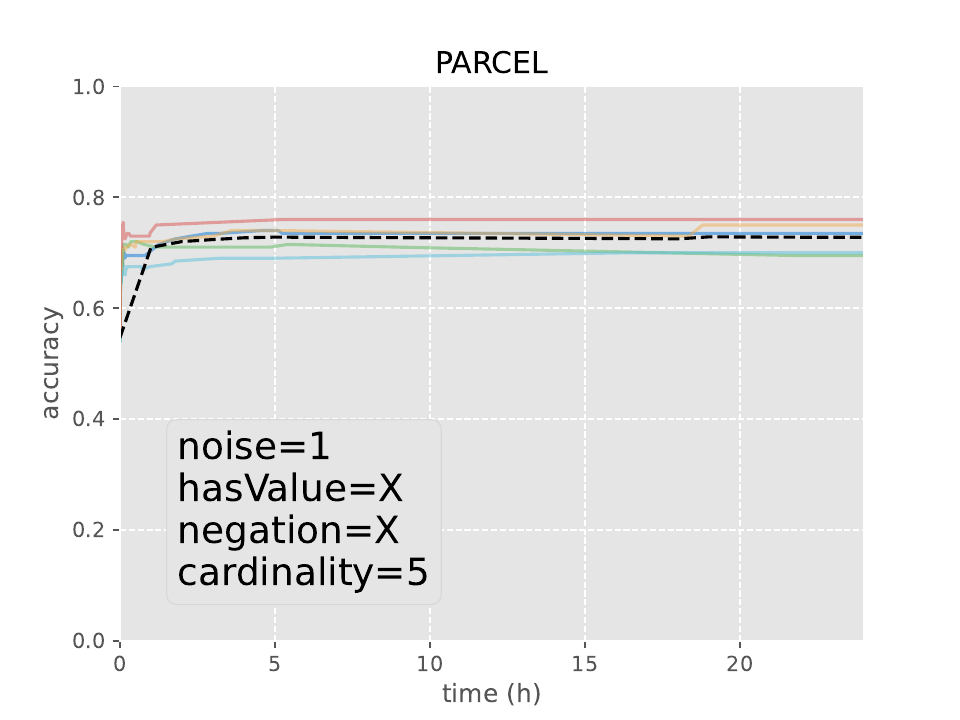}
  \caption{}
\end{subfigure}
\caption{Learning progression of the final configurations on \texttt{dataset\_8\_1000.owl} for PARCEL.}
\label{fig:caliblearning1}
\end{figure}
\begin{figure}[htbp]
\begin{subfigure}{.5\textwidth}
  \centering
  \includegraphics[width=.9\linewidth]{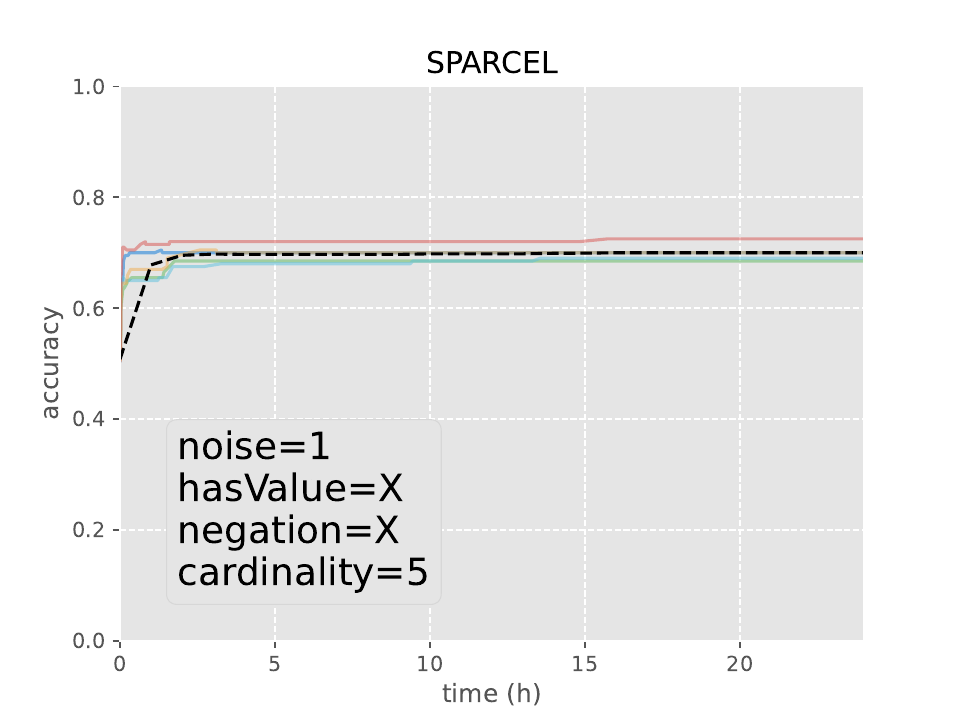}  
  \caption{}
\end{subfigure}\hfil
\begin{subfigure}{.5\textwidth}
  \centering
  \includegraphics[width=.9\linewidth]{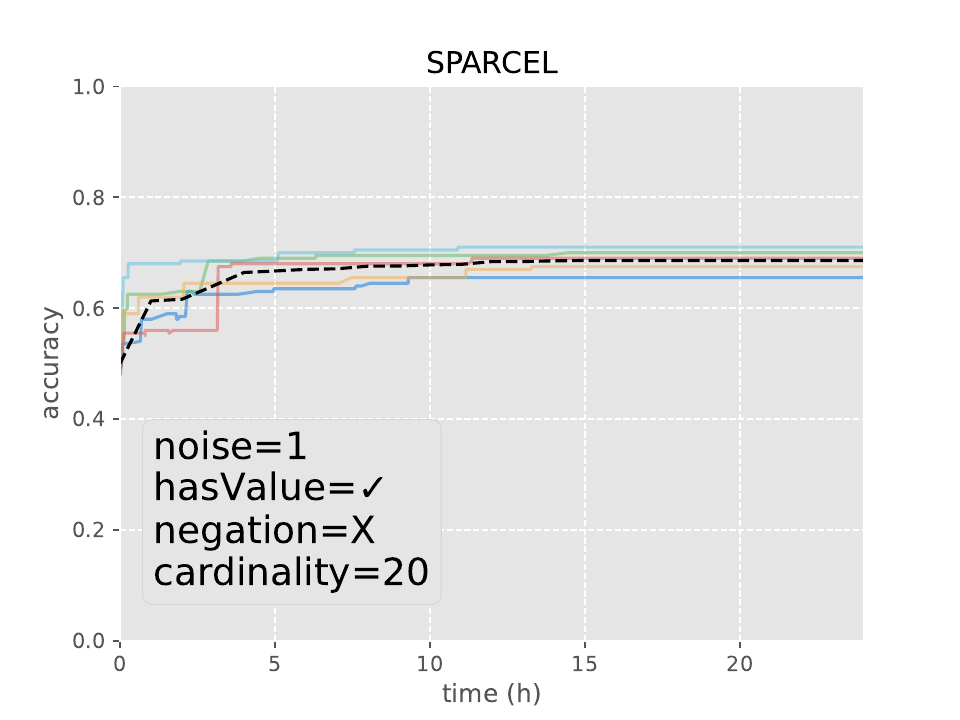}  
  \caption{}
\end{subfigure}

\leavevmode

\begin{subfigure}{.5\textwidth}
  \centering
  \includegraphics[width=.9\linewidth]{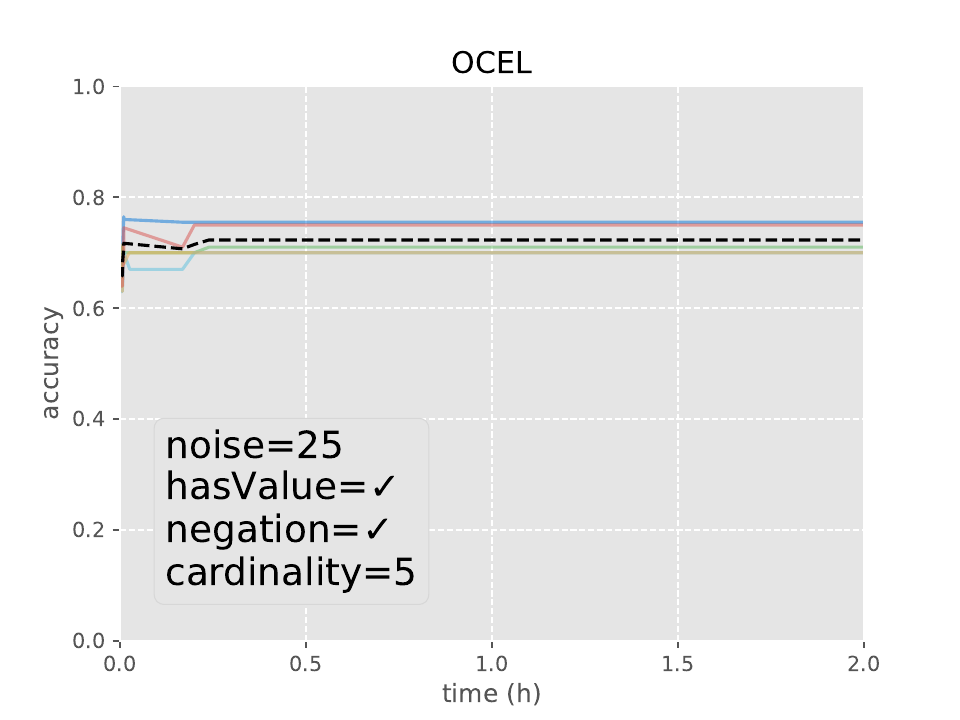}  
  \caption{}
\end{subfigure}\hfil
\begin{subfigure}{.5\textwidth}
  \centering
  \includegraphics[width=.9\linewidth]{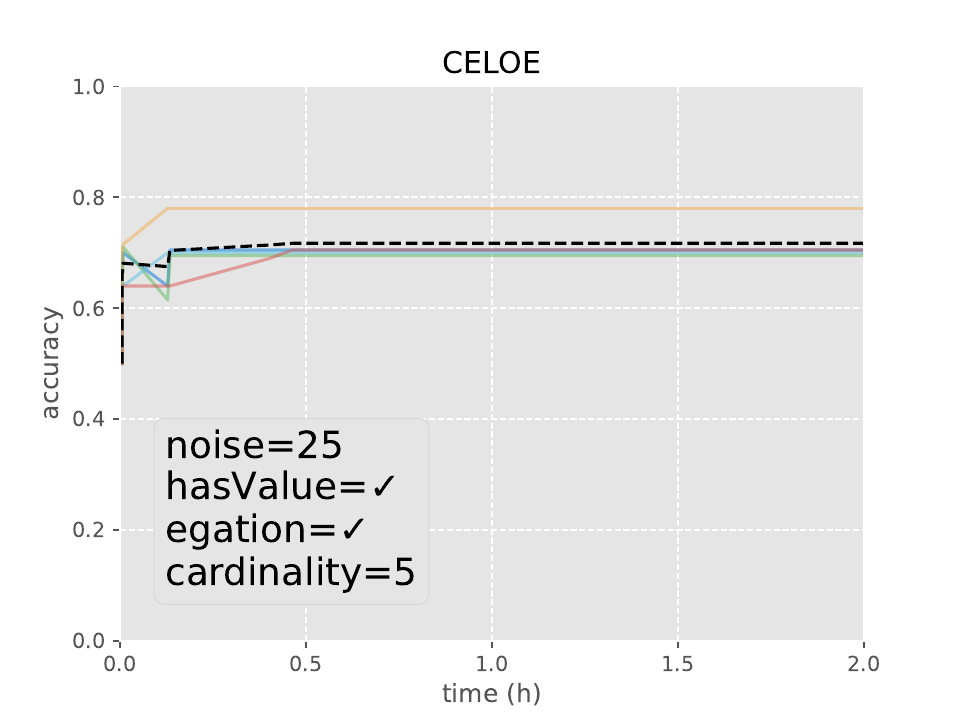}
  \caption{}
\end{subfigure}
\caption{Learning progression of the final configurations on \texttt{dataset\_8\_1000.owl} for SPARCEL, OCEL and CELOE.}
\label{fig:caliblearning2}
\end{figure}

We ran the experiments on two different machines.%
\footnote{
While this setting is not ideal, we had to resort to such a configuration due to the high computational demand of the algorithms, especially under 5-fold cross-validation, supporting which was more important to us.
The differences between the working and running times of the parallel and non-parallel algorithms are so large that an exact match of computing power is not that relevant. It was more important to allow enough computation time in each run, which we assured, as seen in Figures~\ref{fig:caliblearning1} and~\ref{fig:caliblearning2}.}
The first one (12-core AMD Ryzen~9 5900X CPU, 64~GB RAM, Ubuntu 20.04-1) was used for all experiments with the parallel algorithms (PARCEL and SPARCEL), while all experiments with the non-parallel algorithms (OCEL and CELOE) were performed on the second machine (18-core Intel Core i9-10980XE CPU, 256~GB RAM, Debian 5.10.140-1).
%


The training time was measured using the OS-reported \textit{user time} -- the amount of time the CPU spent executing user-space code -- %
which is not dependent on variables such as system load or multitasking delays.

The training user time for the parallel PARCEL and SPARCEL algorithms was established at 24~hours. We configured the algorithms with 12~working threads, so one round of training took approx.\ 2~hours of real-world time. Figures~\ref{fig:caliblearning1} and~\ref{fig:caliblearning2} show that this proved sufficient for the 1\,k datasets, as the training slowed down in all cases before the allocated time elapsed.
To allow better comparison with the non-parallel OCEL and CELOE algorithms, the training time for these algorithms was set to 2~hours of user time. The non-parallel algorithms run in a single execution thread; hence, the user time is in this scenario almost equal to the real-world time.  

\subsection{Calibration}

We have calibrated four parameters:
noise, the use of the \texttt{hasValue} and negation constructors
in generated descriptions, and the cardinality limit, all explained in more detail below.
Note that there are also additional parameters available to calibrate. Specifically, DL-Learner and its $\rho$ operator can refine concepts with data property restrictions. For the purposes of this work, we omitted them; we plan to investigate them in future work.

\paragraph{Noise}
Noise is a common parameter of concept-learning algorithms.
It specifies the minimal acceptable accuracy
of the learned class expression ($a := 1 - \mathit{noise}/100$)
on the training set.
This parameter is either used as the termination criterion,
stopping the training process when the training accuracy reaches~$a$,
or it can be used as a part of the algorithm or heuristics.
By default, this parameter is set to~5.

Noise was the first parameter whose value we calibrated
as it is highly important for the success of the learning process.
During its calibration, we enabled both the \texttt{hasValue}
and negation constructors,
and kept the cardinality limit at its default value of~5.
The noise parameter is used differently by the different algorithms,
which affected the values tested during the calibration,
as well as the results.

\looseness=-1
The PARCEL algorithm focuses on finding partial definitions
that are correct (no negative examples covered)
and the noise parameter controls the percentage of negative examples
the algorithm is allowed to misclassify during the training phase.
We tested four different noise values: 0, 1, 2, and~3.
Higher noise values sped up the training
but, obviously, also increased the FP rates.
We selected 0 and~1 as the best values.
Although noise~0 achieved quite a low F1 score,
we decided to test this setting with other calibration parameters
since it achieved a relatively low FP rate compared to other values.

The SPARCEL algorithm uses the noise parameter solely as the termination criterion
in the way explained above.
We thus just used the constant value of~1
throughout the experiments with SPARCEL,
in order to prevent early termination.

The OCEL and CELOE algorithms use the noise parameter
as both the termination criterion and as a part of their heuristics.
In the latter use, newly refined concepts
that cover less than $\lvert E\rvert \cdot \mathit{noise}/100$ positive examples
are discarded from the search tree.
We progressively increased the noise parameter for both algorithms from 5 up to~25
(five different values in total).
We stopped at 25 since the training accuracy was reaching 0.75,
getting close to the termination criterion.
We observed that with increased noise values,
the accuracy and the precision increased too,
while the recall and the FP rate gradually decreased.
Both algorithms achieved the best results with the noise value of~25.
CELOE achieved a higher F1 score compared to OCEL, although with a higher FP rate.

\paragraph{The \texttt{hasValue} and negation constructors}
Enabling the \texttt{hasValue} constructor allows concept-learning algorithms
(or, more specifically, the $\rho$ refinement operator)
to generate descriptions of the form $\exists r . \{a\}$
with $r$~an object property and $a$~an individual
(also called an individual value restriction).
Similarly, enabling the negation constructor
allows generating concepts of the form of $\neg C$ (the complement of~$C$).
Both settings increase the expressive power
of the generated class expressions
but also enlarge the search space the algorithm explores.
Moreover, the presence of complements in class expressions
may lead to more computationally demanding instance checking.

Testing different combinations
of the \texttt{hasValue} and negation constructors
was the second step in our calibration process.
With the best noise parameters from the initial calibration round,
in which both of these constructors were enabled,
we attempted to find out how disabling
one or both of them
(and the associated reduction of the expressivity and the search space)
affects the performance metrics.

We found out that the parallel algorithms (PARCEL and SPARCEL)
performed best with the \texttt{hasValue} constructor enabled
and negation disabled.
This setting increased the F1 score of PARCEL with noise~0
from 0.43 to 0.71,
and the F1 score of SPARCEL from 0.54 to 0.77.
Disabling both parameters also improved the F1 score
(although not as significantly)
together with a~significant FP-rate drop
for both PARCEL with noise~0 and SPARCEL.
We, therefore, selected both of these parameter combinations
(\texttt{hasValue} enabled with negation disabled, and both disabled)
for further calibration with the two parallel algorithms.
Interestingly, the non-parallel algorithms (OCEL and CELOE) showed
insignificant changes in metrics
for all constructors combinations except the DL-Learner's default
(\texttt{hasValue} disabled and negation enabled).

\paragraph{Cardinality limit}\looseness=-1
The cardinality limit parameter
controls the maximal cardinality~$n$ for which the refinement operator
can generate concepts with qualified cardinality restrictions
${\geq}n\,r . C$, ${\leq}n\,r . C$, and ${=}n\,r . C$,
where $r$~is an object property and $C$~is a description.
If the parameter is~0,  no cardinality restrictions are generated.
Increasing the cardinality limit, again, increases the expressivity
at the expense of enlarging the search space.
DL-Learner's default cardinality limit of~5
was used in the previous calibration steps.


The cardinality limit was the last parameter to be calibrated.
For all parameter combinations selected in the second calibration round,
we were progressively increasing the limit by~10 from~0 up to~30.
Compared to the baseline from the previous round,
only insignificant improvements were achieved in most cases.
For example, the F1 score increased from 0.71 to 0.73
for PARCEL (with noise 0, \texttt{hasValue} enabled, negation disabled)
when the cardinality limit was set to~20.
With non-parallel algorithms, the default value proved sufficient,
as we saw negligible or no F1 improvement.
However, disabling cardinality restrictions entirely
caused a drop in all metrics for almost every setting
with parallel algorithms.
For instance,
the F1 score
of PARCEL with noise~1, \texttt{hasValue} enabled, and negation disabled
declined from 0.75 to 0.69.
With \texttt{hasValue} disabled as well,
the F1 score dropped sharply from 0.68 to 0.32.

\subsection{Results} \label{sec:results}

\paragraph{Accuracy on the calibration dataset}
The calibration was performed on the \texttt{dataset\_\allowbreak 8\_\allowbreak 1000.owl} dataset. Afterwards, we selected eight final configurations, where we included four different configurations for PARCEL, two for SPARCEL and one for OCEL and CELOE. These final configurations, together with their respective results, can be seen in Table \ref{tab:results1}. On the calibration dataset, we consider the parallel algorithms the most successful, in the configurations with \texttt{hasValue} enabled and negation disabled (SPARCEL having F1~= 0.77, PARCEL with noise 0 having F1~= 0.76).
The non-parallel algorithms achieved slightly lower F1 (CELOE was the more successful F1~= 0.72). The learning progression on the above-mentioned eight settings can be seen in Figure \ref{fig:caliblearning1} and Figure \ref{fig:caliblearning2}.

\def\HIDECOL#1\ENDHIDECOL{}%
\newcolumntype{H}[1]{>{\HIDECOL}{#1}<{\ENDHIDECOL}@{}}
\begin{table}[tbp]
    \caption{Results of the calibration process on \texttt{dataset\_8\_1000.owl}.
        Settings format: 
        algorithm(noise/\texttt{hasValue}/negation/cardinality limit).}
    \label{tab:results1}
    \scriptsize\centering\setlength{\tabcolsep}{0.5em}%
    \begin{tabular}{ l c c c H{c} c c }
        \toprule
        {Setting} & {Accuracy} & {Precision} & {Recall} & {Specificity} & {FP rate} & {F1} \\
        \midrule
        \texttt{PARCEL(0/\cmark/\xmark/20)} & 0.76 $\pm$ 0.03 & 0.85 $\pm$ 0.06 & 0.64 $\pm$ 0.04 & 0.88 $\pm$ 0.06 & 0.11 $\pm$ 0.06 & 0.73 $\pm$ 0.03 \\
        \texttt{PARCEL(0/\xmark/\xmark/5)} & 0.70 $\pm$ 0.02 & 0.91 $\pm$ 0.01 & 0.45 $\pm$ 0.06 & 0.95 $\pm$ 0.01 & 0.04 $\pm$ 0.01 & 0.60 $\pm$ 0.05 \\
        \texttt{PARCEL(1/\cmark/\xmark/10)}  & 0.76 $\pm$ 0.04 & 0.76 $\pm$ 0.07 & 0.76 $\pm$ 0.04 & 0.75 $\pm$ 0.09 & 0.24 $\pm$ 0.09 & 0.76 $\pm$ 0.03 \\
        \texttt{PARCEL(1/\xmark/\xmark/5)}  & 0.72 $\pm$ 0.02 & 0.82 $\pm$ 0.04 & 0.58 $\pm$ 0.06 & 0.87 $\pm$ 0.03 & 0.12 $\pm$ 0.03 & 0.68 $\pm$ 0.04 \\ 
        \texttt{SPARCEL(-/\cmark/\xmark/20)}  & 0.77 $\pm$ 0.03 & 0.76 $\pm$ 0.03 & 0.79 $\pm$ 0.02 & 0.75 $\pm$ 0.04 & 0.24 $\pm$ 0.04 & 0.77 $\pm$ 0.02 \\ 
        \texttt{SPARCEL(-/\xmark/\xmark/5)}  & 0.70 $\pm$ 0.02 & 0.91 $\pm$ 0.03 & 0.46 $\pm$ 0.06 & 0.95 $\pm$ 0.02 & 0.04 $\pm$ 0.02 & 0.61 $\pm$ 0.05 \\ 
        \texttt{OCEL(25/\cmark/\cmark/5)}  & 0.72 $\pm$ 0.02 & 0.80 $\pm$ 0.03 & 0.59 $\pm$ 0.03 & 0.85 $\pm$ 0.02 & 0.14 $\pm$ 0.02 & 0.68 $\pm$ 0.03 \\ 
        \texttt{CELOE(25/\cmark/\cmark/5)}  & 0.71 $\pm$ 0.03 & 0.70 $\pm$ 0.03 & 0.74 $\pm$ 0.03 & 0.69 $\pm$ 0.04 & 0.30 $\pm$ 0.04 & 0.72 $\pm$ 0.03 \\ 
        \bottomrule
    \end{tabular}
\end{table}

Notably, the most successful configurations showed a higher F1 but jointly also a quite high FP rate. On the other hand, the final configurations with disabled \texttt{hasValue} and negation, showed a much lower FP rate (0.04 in two cases) but a lower~F1 as well (0.60 and 0.61). In these cases, the learned expressions covered a smaller part of the malware sample but with a higher precision. Such expressions are interesting because learning partial characterizations covering a meaningful portion of the sample precisely is also valuable. They may be indicative of some particular type of malware. Multiple such expressions, even learned on different fractions of the dataset, may potentially be combined using disjunction.

\begin{table}[tbp]
   \caption{Average metrics on five cross-validation datasets.}
    \label{tab:results2}
    \scriptsize\centering\setlength{\tabcolsep}{0.5em}%
    \begin{tabular}{ l c c c H{c} c c }
        \toprule
        {Setting} & {Accuracy} & {Precision} & {Recall} & {Specificity} & {FP rate} & {F1} \\
        \midrule
        \texttt{PARCEL(0/\cmark/\xmark/20)} & 0.68 $\pm$ 0.01 & 0.80 $\pm$ 0.02 & 0.49 $\pm$ 0.03 & 0.87 $\pm$ 0.01 & 0.12 $\pm$ 0.01 & 0.60 $\pm$ 0.03 \\
        \texttt{PARCEL(0/\xmark/\xmark/5)} & 0.62 $\pm$ 0.04 & 0.90 $\pm$ 0.02 & 0.29 $\pm$ 0.09 & 0.96 $\pm$ 0.00 & 0.03 $\pm$ 0.00 & 0.43 $\pm$ 0.12 \\
        \texttt{PARCEL(1/\cmark/\xmark/10)}  & 0.72 $\pm$ 0.01 & 0.71 $\pm$ 0.01 & 0.72 $\pm$ 0.02 & 0.71 $\pm$ 0.01 & 0.28 $\pm$ 0.01 & 0.72 $\pm$ 0.01 \\
        \texttt{PARCEL(1/\xmark/\xmark/5)}  & 0.70 $\pm$ 0.02 & 0.81 $\pm$ 0.01 & 0.52 $\pm$ 0.04 & 0.87 $\pm$ 0.00 & 0.12 $\pm$ 0.00 & 0.63 $\pm$ 0.04 \\ 
        \texttt{SPARCEL(1/\cmark/\xmark/20)}  & 0.72 $\pm$ 0.01 & 0.72 $\pm$ 0.00 & 0.73 $\pm$ 0.02 & 0.72 $\pm$ 0.00 & 0.27 $\pm$ 0.00 & 0.72 $\pm$ 0.01 \\ 
        \texttt{SPARCEL(1/\xmark/\xmark/5)}  & 0.64 $\pm$ 0.03 & 0.88 $\pm$ 0.04 & 0.33 $\pm$ 0.06 & 0.95 $\pm$ 0.01 & 0.04 $\pm$ 0.01 & 0.48 $\pm$ 0.08 \\ 
        \texttt{OCEL(25/\cmark/\cmark/5)}  & 0.69 $\pm$ 0.01 & 0.68 $\pm$ 0.05 & 0.74 $\pm$ 0.10 & 0.64 $\pm$ 0.12  & 0.35 $\pm$ 0.12 & 0.70 $\pm$ 0.02 \\ 
        \texttt{CELOE(25/\cmark/\cmark/5)}  & 0.68 $\pm$ 0.01 & 0.65 $\pm$ 0.03 & 0.77 $\pm$ 0.05 & 0.59 $\pm$ 0.07 & 0.40 $\pm$ 0.07 & 0.70 $\pm$ 0.01 \\         
        \bottomrule
    \end{tabular}
\end{table}

\paragraph{Accuracy on cross-validation datasets}
After the calibration on one dataset, we tested the final configurations on five other datasets of the same size.
Here, we allowed for a reduction of the training time for the non-parallel algorithms to 1 hour of user time, as the calibration showed that they stopped improving after the initial 0.5--1 hour of training (cf.\ Figure~\ref{fig:caliblearning1} and Figure \ref{fig:caliblearning2}). The averaged metrics for all five datasets per configuration are listed in Table~\ref{tab:results2}.

The results from this cross-validation were quite similar, however, the accuracy and F1 scores were lower in most cases (most significantly in configurations where both \texttt{hasValue} and negation were disabled). Two configurations for the parallel algorithms, SPARCEL and PARCEL with noise set to 1, both with \texttt{hasValue} enabled a negation disabled, performed relatively consistently throughout all of the additional datasets. The FP rate increased in almost all cases, with the exception of the three configurations with the lowest calibration FP rate. Also, the precision remained high in all cases but only at the expense of a lower recall (i.e., coverage).

The non-parallel algorithms performed relatively consistently in terms of F1 score, compared to the calibration dataset, but the FP rate increased considerably in some cases.

\paragraph{Learned concepts}
While the parallel algorithms allow approaching the concept learning problem more efficiently (they learn concept expressions with a higher accuracy and a lower FP rate), the resulting concept expressions consist of a number of disjuncts of a considerable total length, which are hence harder to comprehend. We showcase some of the learned partial disjuncts here.

Disjuncts \eqref{ce1} and~\eqref{ce2} were produced by PARCEL. The former can be interpreted as a PE file with at least two sections with a high entropy and a nonstandard name. This is common in packed malware, i.e., if a PE file contains custom sections with encrypted malware code. 
The latter describes a PE file that has a non-executable entry point and possibly performs an action modifying virtual memory protections, e.g., adding the executable permission (common in packed executables or various process injection techniques). Disjuncts from SPARCEL are combined also with negated components abstracted from negative examples (see Disjunct \eqref{ce3}).
\begin{align}
   &\text{\small$\begin{aligned}
       {\geqslant}2\texttt{has\_section} .
        (&\exists\texttt{has\_section\_feature} .
            \{\texttt{high\_entropy}\} \\[-\jot]
         &\sqcap
          \exists\texttt{has\_section\_feature} .
            \{\texttt{nonstandard\_section\_name}\})
   \end{aligned}\!$}
    \label{ce1}
\\
   &\text{\small$\begin{aligned}
    &\exists\texttt{has\_action} .
        \{\texttt{modify-process-virtual-memory-protection}\} \\[-\jot]
    &\sqcap 
    \exists\texttt{has\_file\_feature} .
        \{\texttt{nonexecutable\_entry\_point}\}
    \end{aligned}$}
    \label{ce2}
\\
    &\text{\small$\begin{aligned}
        &\exists\texttt{has\_file\_feature} .
            \{\texttt{low\_imports\_count}\}
          \sqcap
          \exists \texttt{has\_file\_feature} . \{ \texttt{tls} \}
            \!\!\\[-\jot]
        &\sqcap 
            \neg \bigl(
                \exists \texttt{has\_action} .
                    \{ \texttt{close-registry-key} \} \\[-\jot]
        &\phantom{{}\sqcap\lnot(}
                \sqcap \exists \texttt{has\_file\_feature} .
                        \{ \texttt{url-strings} \} \\[-\jot]
        &\phantom{{}\sqcap\lnot(}
                \sqcap \dotsb
            \bigr)
    \end{aligned}$}
    \label{ce3}
\end{align}

Compared to the parallel algorithms, OCEL and CELOE are searching for a single class expression, which tends to be shorter and easier to comprehend. Their expressions, however, achieved a slightly lower accuracy and a higher FP rate in our case. Also, the process of training OCEL and CELOE was more difficult.

Class expressions \eqref{ce4} and~\eqref{ce5} were acquired from OCEL and CELOE, respectively. The former likely indicates a packed executable, while the latter detects the presence of a nonstandard number of MZ headers ($> 1$), indicating multiple embedded executables.
\begin{align}
    &\text{\small$\begin{aligned}
    &\texttt{ExecutableFile} \\[-\jot]
    &\begin{aligned}{}\sqcap
    \exists \texttt{has\_section} .
        \exists \texttt{has\_section\_feature} .
            (&\texttt{HighEntropy} \\[-\jot]
             &\sqcup \texttt{WriteExecuteSection})
    \end{aligned}
    \end{aligned}$}
    \label{ce4}
\\
    &\text{\small$
    \texttt{ExecutableFile} \sqcap
    \exists \texttt{has\_file\_feature} . \{ \texttt{nonstandard\_mz} \}
    $}
    \label{ce5}
\end{align}

From the viewpoint of a malware analyst, such learned expressions are plausible descriptions of possible malware instances, even if -- thus far -- they are still rather generic. This is caused by the limitations of our preliminary experiments: the smaller datasets size and their randomly selected structure. (See discussion in Section~\ref{sec:usecase:discussion} on how to potentially overcome this issue.) 

However, the most important result, which underlines the value of the overall approach, lies in the high interpretability of the learned expressions. Each learned expression enables the expert users (i.e. security professionals) to clearly understand the features by which data instances are classified as malware or benign and also \emph{the exact logic of this decision} (i.e., it captures also the exact inter-relation of the features in logical terms). We believe that this brings a strong level of interpretability, novel in this domain, which is immediately apparent and easy to understand for the users. To further corroborate this claim we have conducted a user survey on five independent security analysts (external to our team of co-authors). We have presented each participant the learned expressions \eqref{ce1}--\eqref{ce5}%
\footnote{The participants were informed about the fact that the formulas \eqref{ce1}--\eqref{ce3} resulting from the parallel algorithms are just selected disjuncts from the overall learned expression, and that it means that they are partially responsible for the positive classification of a respective part of the sample.}
(both in the DL syntax as printed above and transliterated to English). With each expression, the participants answered the following questions:

\begin{quote}\it
\begin{enumerate}
\item[Q1.] Is the given justification indicative of a sample possibly being malware?
\item[Q2.] Does the reading of the formula or its transliteration help you to understand why the system classifies the samples as malware/benign?
\item[Q3.] Is an explanation/justification of this form useful compared to black-box malware detection methods or compared to post-hoc explainers such as LIME or SHAP?
\end{enumerate}    
\end{quote}

\begin{wrapfigure}{r}{0.4\textwidth}\vspace*{-4ex}
    \includegraphics[width=\linewidth]{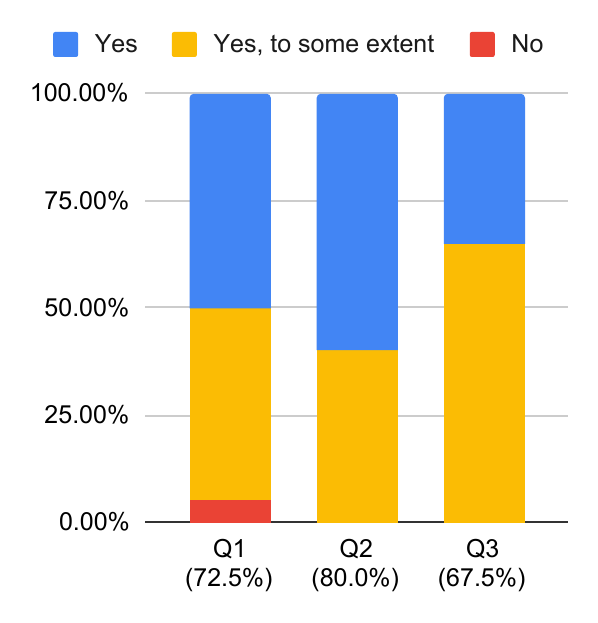}
    \caption{User study results}\vspace*{-4ex}
    \label{fig:chart_user-study}
\end{wrapfigure}

\noindent
The answers were selected from the scale:
\begin{quote}
\begin{enumerate}
\item Yes
\item Yes, to some extent
\item No
\end{enumerate}    
\end{quote}

The average answers over the five expressions and over the five participants are shown in the chart in Figure~\ref{fig:chart_user-study}.
We have computed also the average weighted score for each question (given in brackets), where the first, second, and third answer were assigned the weight of 100\%, 50\%, and 0\%, respectively. We observe that the overall the first two answers largely dominate against the clearly dismissive third answer. Particularly encouraging for us are the results for Q2, the affirmative answer ``Yes'' prevails and the overall weighted score is 80\% -- this is supportive of our claim above.

As we can see, the explanations obtained through concept learning achieve better interpretability compared to those from LIME (Figure \ref{fig:lime_exp}) and SHAP (Figures \ref{fig:shap_boxplot} and \ref{fig:shap_exp}). While post-hoc methods capture the contributions of individual features, class expressions also capture logical relationships between features, which can be useful for experts. This fact was also confirmed by our user survey.

\color{black}

\subsection{Discussion}
\label{sec:usecase:discussion}
In the case study, we evaluated four concept-learning algorithms available in DL-Learner. This case study is preliminary, in the sense it serves mainly to prove the basic feasibility of this approach. There is space to improve the results, as we discuss here.

The lessons we learned immediately from our experiments:
\begin{itemize}
    \item Learning even on small fractions of the dataset (1\,k = 0.125\,\% of the annotated EMBER sample) produced interesting concept expressions, meaningful from the malware-analysis point of view. As discussed and showcased in the previous section, 
    the learned expressions showed good interpretability from the malware expert's point of view. The experts understood the expressions easily and were able to immediately link them with the respective combination of traits in the sample indicative of potential malware behaviour.
    \item The parallel algorithms achieved a higher accuracy and a lower FP rate, but they found less comprehensible concept expressions (disjunctive and rather long). In contrast, the non-parallel algorithms found concise and comprehensible expressions but with a slightly lower accuracy and a higher FP rate. They required more memory even during shorter runs.
    \item Concept learning seems to be applicable, but it is rarher time-consuming even on the smallest fractional datasets with 1\,k of examples. It is not feasible to run DL-Learner on current off-the-shelf consumer hardware on full EMBER-sized datasets.
\end{itemize}

These are but the first findings we were able to obtain, and they open a number of questions to be tackled in the future:
\begin{itemize}
    \item The achieved accuracy is so far lower compared to state-of-the-art ML methods (almost 90\,\% on EMBER with more interpretable features and over 90\,\% using all features \citep{oyama19useful}, but approaching 100\,\% in some experiments with other datasets \citep{shaukat2020survey}).
    This is partly due to the exclusion of non-interpretable features (e.g.\ byte histograms) which are known to significantly improve the accuracy \cite{oyama19useful}. Still, there is space for improvement, e.g., by processing larger datasets.
    
    \item Handling larger datasets is particularly challenging. A 10\,k dataset may be processed by increasing the computational time to the order of days (at least with the less memory-demanding parallel algorithms), but moving to 50--100\,k is hardly as simple. A possible approach may be to partition the inputs (e.g., by a deeper domain analysis or by clustering), and then evaluate and recombine the learned expressions.
    Improving the quality of the datasets by sample balancing and pre-filtering may enable to improve accuracies of the learned expressions, at least to certain extent, even without necessarily increasing the dataset size.
    Another approach would be to learn less expressive concepts in tractable DLs by systems such as SPELL \citep{SPELL}. This will require jointly investigating if the learned expressions are sufficiently expressive to discriminate the sample, as well as their usefulness for malware experts.
    \item Malware datasets contain a significant number of numeric values. Applying fuzzy concept learners \citep{straccia2015pfoil,cardillo2022fuzzyOWLboost,cardillo2023PN-OWL} 
    or DL-Learner's data property-handling capabilities on such data might help to improve the accuracy.
\end{itemize}

\section{Related works}
\label{sec:related}

Ontologies found their use in the field of computer security as knowledge bases of attack patterns, malicious code, or various security incidents \cite{bromander2016semantic, mavroeidis2017cyber}. Among some well-known ontologies, we can mention the Unified Cybersecurity Ontology (UCO), which integrates several ontologies and sharing standards from the field of computer security \cite{syed2016uco}.  Ulicny et al.\ \cite{ulicny2014inference} presented the idea of automatically generating an ontology from known sharing standards that serve to exchange information regarding security incidents. They focused on the Structured Threat Information Expression (STIX) standard, where together with the automatic generation of the ontology, they also presented the improvement of the analytical process in response to security incidents through automatic reasoning. Carvalho et al.\ \cite{Carvalho2019Semantic} presented a solution that helps investigators of various security incidents by searching connections using SPARQL queries. Xia et al.\ \cite{XiaDJZ17} proposed using the data-mining Apriori algorithm for mining rules from ontologies. In their work, the authors presented an ontology focused on dynamic properties. A similar work was also proposed by Ding et al. \cite{DingWZ19}, where they focused on individual malware families. They achieved 96\% accuracy with mined rules (although on a relatively small dataset).

Numerous malware datasets were collected and made available to researchers (see Section~\ref{sec::othersources} for a comparison). To our best knowledge, apart from our work, there were no prior attempts to facilitate ontologies to enrich malware datasets with the semantics required to employ knowledge-enabled AI tools such as concept learning to characterize malware for the sake of explainability. Closest to our efforts is perhaps the Malicious Behavior Ontology (MBO), which describes malware \cite{gregio2016ontology}. The primary objective of this ontology is to represent various malicious dynamic properties, such as anti-analysis, system information stealing, or exploitation. It may however be useful for the future goal of extending the PE Malware Ontology for the needs of dynamic malware analysis (as we discuss below). The MALOnt ontology is also a knowledge base that describes malware \cite{rastogi2020malont}, but its main focus is threat intelligence (i.e., attack patterns, IoCs, etc.). However, it should be noted that none of the mentioned ontologies are directly compatible with state-of-the-art static malware datasets, such as EMBER, \ac{SOREL}, or BODMAS. Also, they does not follow the requirements for this type of application, as given by Balogh \cite{balogh2021}.

Galli et al.\ \cite{galli2024explainability} present a XAI framework for Behavioral Malware Detection (BMD), aiming to evaluate the effectiveness, strengths, and weaknesses of four distinct XAI methods. They compare explanations provided by Local Interpretable Model-agnostic Explanations (LIME), SHapley Additive exPlanation (SHAP), Layer-wise Relevance Propagation (LRP), and attention mechanisms when explaining predictions from Long-Short Term Memory (LSTM) and Gated Recurrent Unit (GRU) models. 
But the work focuses only on a model that processes API call sequences as input and outputs malware class predictions. They use three datasets with API calls Mal-API-20193 \cite{catak2020deep}, API Call Sequences4 \cite{de2023behavioral} and Alibaba Cloud Malware5\footnote{https://tianchi.aliyun.com/getStart/information.htm?raceId=231694}  datasets. In discussion section they highlight that further research is needed in the XAI field to achieve the goals that XAI techniques can bring in malware detection area.
Similarly Manthena et al.\ \cite{manthena2023analyzing}, applied SVM Linear, SVM-RBF, RF, FFNN, and CNN models on an online malware dataset to evaluate which model suited the best to detect malware, and three SHAP methods such as KernelSHAP, TreeSHAP, and DeepSHAP were used to explain the feature contributions. 
Their dataset was collected by performing an experiment on a cloud testbed environment. They used for experiments only 113 malware samples from VirusTotal which includes samples from various malware categories such as Trojan, DDoS, Virus, Backdoor, DoS, Constructor, Exploit, Worm, HackTool, Net-Worm, Trojan-DDoS, and VirTool.

\section{Conclusions}
\label{sec:conclusions}

The \emph{PE Malware Ontology}, the main contribution of this paper, is 
intended to serve as an interoperable semantic representation format that can be used to publish datasets resulting from static malware analysis of PE files (i.e., the executables and libraries of the Windows platform). Such a unified representation will particularly facilitate (i)~the syndication and importing of data from various sources to be processed, (ii)~the application of some of the more advanced AI tools, including neuro-symbolic tools \cite{NSAI}, which may require semantic representation, and (iii)~better interoperability and comparison of the achieved results.

The ontology is partly based on the \acs{EMBER} dataset but does not copy the EMBER data properties one-to-one. Instead, it relies on expert knowledge and standard nomenclature to improve interpretability and reusability. To this end, it also incorporates the MAEC standard for encoding software actions. Being based on EMBER, the ontology is currently suited especially for data resulting from static malware analysis. 
We hope that the ontology could serve as the first step towards a more widely accepted model of a semantic schema specifying what data should actually be included in these datasets -- also considering its future gradual extensions, as needed. 

Together with the ontology, we have also published datasets of different sizes derived from EMBER, arguably\footnoteref{footnote1}
currently the most popular dataset in malware research. However, this approach is applicable also to other datasets gathering data from static analysis (e.g.\ \ac{SOREL} and BODMAS). To demonstrate this, we have also extracted and published a smaller number of datasets from SoReL. 

Finally, we have also reported on a feasibility case study, that applied concept learning using DL-Learner \cite{DL-Learner} on the semantically treated EMBER data. As we were able to demonstrate, thanks to representing the data using our ontology, concept learning may be used to extract meaningful concept expressions characterizing the malware sample in the dataset. As discussed in Section~\ref{sec:results}, such symbolic expressions can be straightforwardly transliterated into human language and they showed very good interpretability by malware experts. 

Further results achieved by concept learning on the \ac{SOREL} datasets were published separately \cite{dl2023}. But the semantic datasets presented in this work enable to apply also other methods to construct some form of explanations for malware classifiers, where the ontology provides a unifying and mutually intelligible language in which these explanations are expressed.
It is noteworthy that, since these datasets were first published, they have already been used in at least four independent studies applying other methods than we did in the preliminary use case presented herein, including decision-tree learning \cite{MojisKenyeres2023}, fuzzy concept learning \cite{cardillo2023PN-OWL}, and two neuro-symbolic approaches, logic-explained networks \cite{tailored-LEN} and knowledge-base embedding \cite{Trizna_EMBER_embeddings}.

This approach may be used on collected pre-annotated datasets, as in this study, where the method yields a primary, inherently interpretable classifier; but it can also be used ex-post for processing the classification results obtained by one of the popular black-box ML classification models, empowering them with ex-post explainability enabling deeper insights than e.g.\ the popular feature importance methods \cite{LIME,SHAP}.

In the future, we plan to conduct a more comprehensive analysis of the EMBER, SoReL, and BODMAS datasets with the aim to process a more significant fraction of the data applying concept learners, but also other relevant methods (decision-tree learning \cite{MojisKenyeres2023}, inductive logic programming \cite{HocquetteNJC24}, knowledge-base embedding \cite{TRANSE}, neural-network activity pattern recognition \cite{deSousaRibeiro2021}, logic explained networks \cite{LEN}, graph neural networks \cite{GNN} etc.).

We especially hope that committing more computational resources, applying pre-processing methods including pre-filtering and balancing of data, and exploiting possible data breakdown strategies (e.g., by a deeper domain analysis or by clustering) which may enable to process smaller datasets separately and then recombine the outputs, will help us to improve generalization and the resulting precision of the learned expressions. 

While we also presented a very preliminary user study as part of our use case, further research is needed to answer the fundamental question of what properties the extracted explanations should have and how they should be presented to be the most useful to the expert users they are intended to. For such research, a larger pool of representative learned characteristic expressions with different levels of expressivity should be systematically gathered, ideally obtained by multiple distinct methods, and they should be thoroughly evaluated in user studies comparing them based on size, complexity and different modes of presentation. Such research results are needed to guide the selection and configuration of the most proper methods to obtain explanations that are not only precise but at the same time also well-understandable and thus actually useful to the users.

Finally, being based on EMBER and \ac{SOREL}, the ontology thus far covers only data resulting from static malware analysis; however, it is of interest to extend it towards dynamic malware analysis \cite{sikorski2012practical} as well. Ontologies, in general, allow for good extensibility \cite{modular-ontologies} and integration. Indeed, in the future, the PE Malware Ontology may be extended to support other relevant data about the sample files, in particular data obtained by dynamic malware analysis.%
\footnote{%
As discussed in Section~\ref{sec:actions}, actions performed by processes when running the sample can already be individually represented. The processes and system or network objects affected by actions can be represented by extending the ontology with the respective classes. Additional properties can relate samples to processes, processes to actions, and actions to the affected objects, akin to how these relationships are represented in the JSON format in the MAEC standard \cite{maec}.} 
This would then allow applying the same methodology outlined in this paper to more refined datasets. The nature and amount of details covered by the extended ontology that will balance the computational requirements, classification accuracy, and explanation quality is a subject of further research much beyond the scope of this paper. 
\par

\section{Acknowledgement}
The authors would like to thank Claudia d'Amato, Fran\-ce\-sco Giannini, Ján Mojžiš, Umberto Straccia, Alexander Šimko, and Pavol Zajac for consultations and feedback related to this work. This research was sponsored by Slovak Republic under the SRDA grant
APVV-23-0292 (DyMAX) and VEGA grant no.\ 1/0630/25 (eXSec).


\bibliographystyle{elsarticle-num} \bibliography{references}

\end{document}